\documentclass[aps,prl,
reprint,
longbibliography,notitlepage,superscriptaddress]{revtex4-2}

\usepackage{orcidlink}
\usepackage{physics}
\usepackage{graphicx}% Include figure files
\usepackage{dcolumn}% Align table columns on decimal point
\usepackage{lipsum}
\usepackage{float} % in your preamble

\usepackage{amsfonts}

\usepackage[normalem]{ulem}
\usepackage{ragged2e}
\usepackage{flushend}

\usepackage{bm}% bold math
\usepackage{xcolor}
\definecolor{myteal}{HTML}{0F6E6C} % tweak hex as you like

\newcommand{\vphi}{\varphi}
\newcommand{\crat}{\zeta}
\newcommand{\ra}{\rightarrow}

\newcommand{\mc}[1]{\mathcal{#1}}

\newcommand{\mief}{\mathrm{MIE_F}}

\DeclareUnicodeCharacter{2009}{\,} 
\usepackage{hyperref}
\hypersetup{
    colorlinks=true,
    linkcolor=blue,
    citecolor = blue,
    filecolor=magenta,      
    urlcolor=blue,
    pdftitle={MIEdraft}
}

\usepackage{pdfpages} % include appendix
\makeatletter
\AtBeginDocument{\let\LS@rot\@undefined}
\makeatother

\newcommand{\umass}{Department of Physics, University of Massachusetts, Amherst, MA, USA}
\newcommand{\unige}{Department of Theoretical Physics, University of Geneva, 24 quai Ernest-Ansermet, 1211 Gen\`eve, Switzerland}

\begin{document}

%%\preprint{APS/123-QED}
\newcommand{\romain}[1]{\textcolor{red}{(RV): #1}}

\title{Measurement-Induced Entanglement in Conformal Field Theory}% Force line breaks with \\

\author{Kabir Khanna\:\orcidlink{0000-0001-9400-1597}}
\affiliation{\unige}
\affiliation{\umass}

\author{Romain Vasseur\:\orcidlink{0000-0002-4636-4139}}
 %\email{romain.vasseur@unige.ch}
\affiliation{\unige}

% \collaboration{MUSO Collaboration}%\noaffiliation

%  \homepage{http://www.Second.institution.edu/~Charlie.Author}
% \affiliation{
%  Second institution and/or address\\
%  This line break forced% with \\
% }%
% \affiliation{
%  Third institution, the second for Charlie Author
% \affiliation{%
%  Authors' institution and/or address\\
%  This line break forced with \textbackslash\textbackslash
% }%

% \collaboration{CLEO Collaboration}%\noaffiliation

\date{\today}% It is always \today, today,

\begin{abstract}
Local measurements can radically reshape patterns of many-body entanglement, especially in long-range entangled quantum-critical states. Yet, analytical results addressing the effects of measurements on many-body states remain scarce, and measurements are often approximated as forcing specific measurement outcomes. We study measurement-induced entanglement (MIE) in Tomonaga-Luttinger liquids, a broad family of  1+1d quantum critical states described at low energies by compact free boson conformal field theories (CFT). Using a replica-trick to address the randomness of the measurement outcomes, we compute exactly the entanglement induced by measuring the local charge operator for Tomonaga-Luttinger liquids, in very good agreement with matrix-product state calculations. We show that the MIE for physical quantum measurements is fundamentally different from the entanglement induced by forcing measurement outcomes, and has a natural interpretation in terms of Born averaging over conformally-invariant boundary conditions.

\end{abstract}
\maketitle
\textit{Introduction --- } Entanglement is a cornerstone of quantum theory and a powerful diagnostic of quantum phases of matter \cite{PhysRevA.43.2046, srednicki1994chaos, HOLZHEY1994443, Pasquale_Calabrese_2004, kitaev2006topological, levin2006detecting, PhysRevLett.96.181602, RevModPhys.80.517, calabrese2009entanglement, annurev:/content/journals/10.1146/annurev-conmatphys-031214-014726, abanin2019colloquium, PhysRevX.7.031016}. Its most striking features—particularly its nonlocal character—become evident in the presence of measurements, as exemplified by quantum teleportation \cite{PhysRevLett.70.1895, bouwmeester_experimental_1997} and entanglement swapping \cite{PhysRevLett.70.1895, PhysRevLett.71.4287, PhysRevA.57.822, PhysRevLett.80.3891}. These insights have sparked interest in the interplay between measurement and entanglement, particularly with the advent of measurement-based quantum computation (MBQC) \cite{PhysRevLett.86.5188, PhysRevA.68.022312, briegel_measurement-based_2009}, where local measurements drive computation on a resource state. While early work focused on the many-body cluster state~\cite{PhysRevA.68.022312}, later studies demonstrated that entire phases of matter can support universal computation \cite{PhysRevLett.119.010504, PhysRevA.96.012302, PhysRevA.98.022332, PhysRevLett.122.090501,daniel2020computational}. More recent efforts have shown that local measurements can create long-range entanglement~\cite{PhysRevX.14.021040, verresen2021efficiently, PRXQuantum.3.040337, iqbal_non-abelian_2024, foss2023experimental, PhysRevLett.131.200201, chen_nishimori_2025, PhysRevA.107.032215, Chan_2024, PhysRevLett.128.060601, Ippoliti2022solvablemodelofdeep, choi_preparing_2023, PRXQuantum.4.010311, PRXQuantum.4.030322, PhysRevX.14.041051, 10.21468/SciPostPhys.18.3.107, milekhin2024observable}, induce criticality~\cite{PhysRevX.9.031009, PhysRevB.98.205136, PhysRevB.100.134306, PhysRevB.101.104301, PhysRevB.101.104302, hoke_measurement-induced_2023, koh_measurement-induced_2023, annurev:/content/journals/10.1146/annurev-conmatphys-031720-030658, Potter2022} and also alter it~\cite{garratt2023measurements, PhysRevX.13.041042, PhysRevB.108.165120, sun2023new, PhysRevA.94.053615, 10.21468/SciPostPhys.12.1.009, PhysRevX.11.041004}--further broadening quantum-computational applications while deepening our understanding of phases of matter.

The dual role of measurements---as both a key ingredient in quantum computation protocols and a tool for probing quantum phases---motivates a detailed examination of how local measurements reshape entanglement in many-body systems. A central quantity in this context is the measurement-induced entanglement (MIE) \cite{Lin2023probingsign}, which quantifies the long-range entanglement between two distant regions after the rest of the system has been locally measured. The MIE of a region $A$ is defined as the entanglement entropy of $A$ after part of the system has been measured, averaged over \textit{all} measurement outcomes weighted by their Born probabilities. This averaging captures genuine measurement-induced correlations, as opposed to its forced counterpart, labeled as $\rm{MIE}_F$, where one typically post-selects to a single measurement outcome. MIE was first used in the context of localisable entanglement (LE) \cite{PhysRevLett.92.027901}, where it was used to bound two-point correlations. Since then, MIE has proven operationally useful in other settings. For example, a non-zero long-range MIE is been known to be necessary (though not sufficient) for MBQC \cite{briegel_measurement-based_2009}. Recent work has shown that whenever MIE exceeds the pre-measurement mutual information, the resulting wavefunction necessarily develops a sign problem in any product basis—linking MIE to the complexity of simulating quantum phases \cite{hastings_how_2015, Lin2023probingsign}. More broadly, MIE governs the classical memory and quantum resources required for certain tensor-network contractions \cite{PhysRevResearch.3.033002, chertkov_holographic_2022, PhysRevX.15.021059} and upper-bounds strange correlators used to diagnose symmetry-protected topological (SPT) order \cite{PhysRevLett.112.247202}. Beyond these operational roles, studies have found that MIE can exhibit universal behavior \cite{PhysRevB.109.195128}. In particular, in 1+1D systems, the leading long-range contributions to MIE appear universal, whereas in 2+1D, universality emerges only in subleading terms \cite{PhysRevB.109.195128, PhysRevLett.108.240505}. A priori, this is far from obvious given that MIE involves averaging over Born probabilities of all measurement outcomes. The emergence of universality thus prompts a deeper investigation into its origin, and into the extent to which MIE reflects universal features of quantum phases. 

In this letter, we obtain the first analytic result for MIE in a class of long-range-entangled ground states: 1+1D quantum-critical states described by the compact free-boson CFT. A natural strategy to evaluate the MIE is to build on previous calculations for $\mathrm{MIE_F}$ \cite{PhysRevB.92.075108, Rajabpour_2016, najafi_entanglement_2016, PhysRevB.111.155143},  where the measurement outcome is fixed, e.g., to the state $\ket{1010\dots}$, corresponding to a Dirichlet conformal boundary condition. The measured region then acts as a slit/defect on the manifold and the problem can be simplified using the boundary CFT (BCFT) formalism \cite{CARDY1984514, cardy2004boundary, OSHIKAWA1997533}. While these techniques form an essential ingredient in our calculation, the MIE resists such a treatment due to the randomness of the measurement outcomes. We overcome this difficulty by employing the replica trick, owing to its success in addressing measurement-related disorder~\cite{ PhysRevB.101.104301, PhysRevB.101.104302}, along with free-boson specific techniques developed in other contexts~\cite{PhysRevLett.97.050404, PhysRevB.84.195128, PhysRevB.90.045424, Zhou_2016}. Both $\mief$ and MIE are universal, conformally invariant, and we provide closed-form expressions for them. Importantly, we remark that while the MIE shares some of its contributions with $\mathrm{MIE}_\mathrm{F}$, it contains additional contributions that arise from measurement-physics. Our final result has a natural interpretation in terms of ``Born-averaging over  conformal boundary conditions'', which we expect to hold generally for CFTs.

\begin{figure}[t!]
    \centering
    \includegraphics[width=\linewidth]{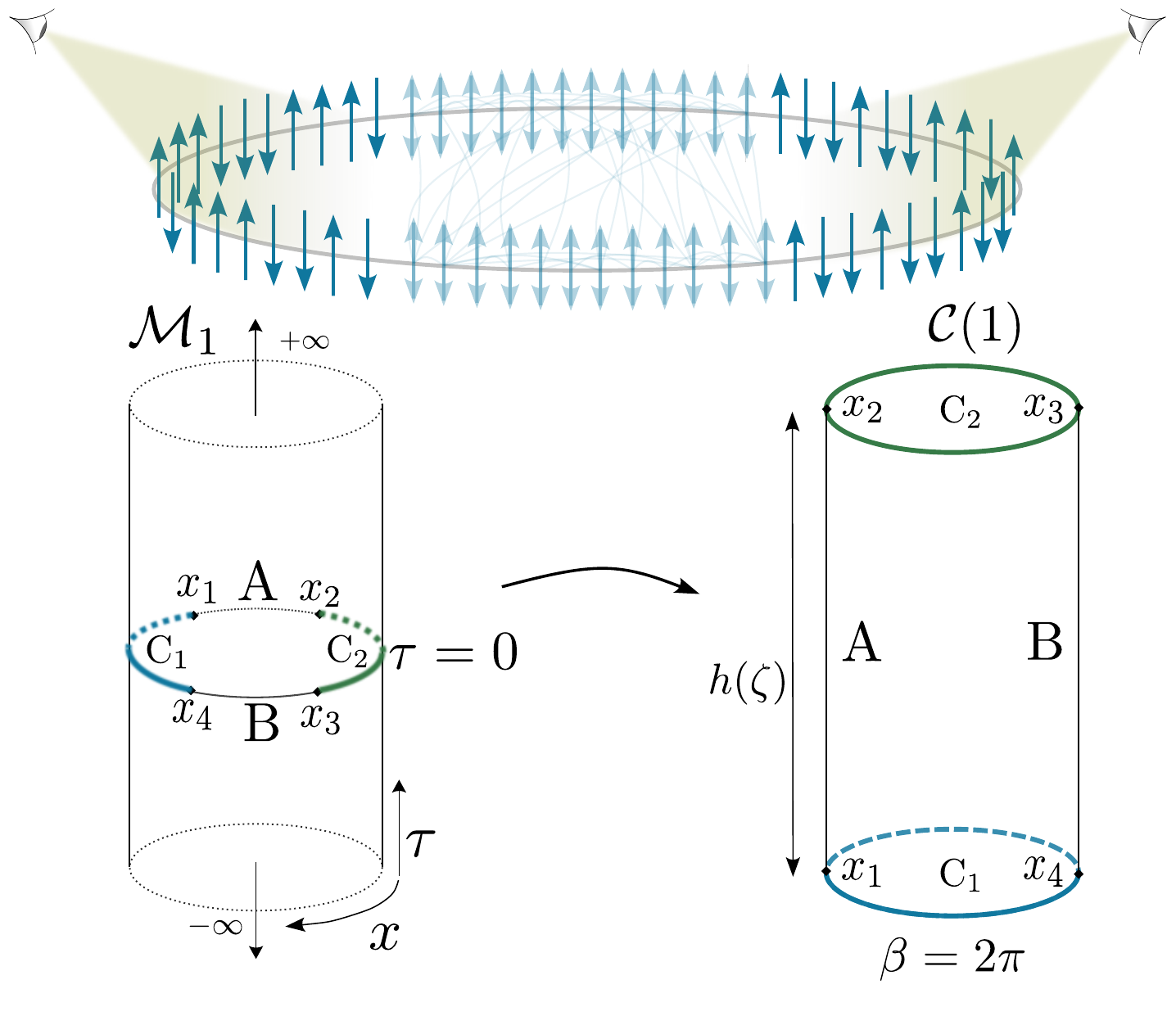}
    \caption{\textbf{Setup. }\textbf{Top}: Schematic illustration of a TLL modeled as a spin-1/2 chain. Dark (light) blue sites denote measured (unmeasured) spins. Blue arcs indicate long-range entanglement between the unmeasured spins. \textbf{Bottom}: Manifolds $\mc{M}_1$ and $\mc{C}(1)$, and the conformal mapping between them. The intervals $\mathrm{A} = [x_1,x_2]$ and $\mathrm{B} = [x_3,x_4]$ on the cylinder $\mc M_1$ denote unmeasured regions, while $\mathrm{C} = [x_1,x_4]\cup[x_2,x_3]= \mathrm{C}_1 \cup \mathrm{C}_2$ denotes the measured region. $\mc C(1)$ has circumference $\beta=2\pi$ and length ${h(\crat)}$ which is a function of the conformal cross ratio $\crat=w_{12}w_{34}w^{-1}_{13}w_{24}^{-1}$, where $w_{ij} = \frac{L}{\pi} \sin(\pi x_{ij}/L)$.}
    \label{setup} \label{Fig1}
\end{figure}

\textit{Setup--- } We begin by considering a broad class of quantum states in 1D whose universal low-energy properties are described by Tomonaga-Luttinger liquids (TLL) \cite{tomonaga_remarks_1950, luttinger_exactly_1963}, including interacting metallic states of fermions or many gapless quantum spin chains for example \cite{giamarchi_quantum_2003, F_D_M_Haldane_1981, PhysRevLett.47.1840}. 
At low energies, the physics of Luttinger liquids is described by a compact free boson CFT with 
Lagrangian density
\begin{equation}
\label{euclideanfreebosonaction}
    {\cal L} = \frac{g}{4\pi } \left( \partial_\mu\vphi\right)^2,
\end{equation}
where $g$ is the Luttinger parameter that characterizes this continuous family of CFTs. Here, the coarse-grained field $\vphi$ acts as a ``counting field" for the $U(1)$ charge, and the microscopic charge operator is given by $ \hat{n}(x) \simeq n_0 + \frac{1}{2\pi}\partial_x {\vphi} + A\cos({\vphi}(x) +2\pi n_0 x) +\dots$ \cite{giamarchi_quantum_2003}, where $A$ is a non-universal constant, $\rho_0$ the background filling fraction, and $\dots$ represent higher-order harmonics that can be ignored at large distances. The field $\vphi$ is a compact variable with unit compactification radius: $\vphi \sim \vphi + 2\pi w $ with $w \in \mathbb{Z}$ the winding number. 

We now define the central quantity of interest in this work: the measurement-induced entanglement (MIE). For concreteness, we begin by placing our TLL on a ring and perform projective measurements of the local charge operator $\hat{n}$ in two disjoint well-separated regions $C = C_1 \cup C_2$ (see figure \ref{setup}), which corresponds to measuring the field $\vphi$ in the field theory limit \cite{garratt2023measurements, nahum2025bayesian}. The measurement outcomes are denoted by ${\bf m}$, and occur with Born probability $p_{\bf m} = {\rm tr} \rho_{\bf m}$ with $\rho_{\bf m}$ the (un-normalized) pure state post-measurement density matrix of the system. The MIE probes the entanglement induced between the remaining, spatially separated regions $A$ and $B$ (see figure \ref{setup}), and is defined as
\begin{equation}
\label{miedef}
    \mathrm{MIE^{}}(A) = \sum_{\bf m}p_{\bf m}S_{\bf m}(A) ,
\end{equation}
where $S_{\bf m}(A) = -\tr(\rho_{{\bf m},A}\log\rho_{{\bf m},A})$ is the Von-Neumann entanglement entropy of region $A$ conditional on the measurement record ${\bf m}$ in region $C$, with $\rho_{{\bf m},A} = \tr_B \rho_{\bf m} / \tr \rho_{\bf m}$. We also define a “forced” (or post-selected) version of the MIE, denoted \(\mathrm{MIE}_{\mathrm{F}}(A,{\bf m_0})\), in which one post-selects a specific uniform outcome \(\bf m_0\)
\begin{equation}
\label{mieforceddef}
    \mathrm{MIE_{F}}(A,{\bf m_0}) = S_{\bf m_0}(A).
\end{equation}
The Rényi versions of the above MIEs naturally follow and are denoted as $\mathrm{MIE}^{(n)}$ and  $\mathrm{MIE_F}^{(n)}$ respectively. 

\textit{Replica trick and path integral ---} Our first step to address the non-linearity of~\eqref{miedef} is to use a double replica trick
\begin{equation}
\label{replicalimit}
        \mathrm{MIE}(A) =  
        \lim_{n\ra 1}\lim_{k \to 0}\frac{1}{1-n} \frac{ d}{{ d} k} \log \left( \frac{\mathcal{Z}_A}{\mathcal{Z}_0} \right),
\end{equation}
where $\mathcal{Z}_A = \sum_{\bf m} p_{\bf m} \left(\tr \rho_{{\bf m},A}^n\right)^k$ and $\mathcal{Z}_0 = \sum_{\bf m} p_{\bf m} \left(\tr\rho_{\bf m}\right)^{nk} = \sum_{\bf m} p_{\bf m}^{nk +1} $, with $n$ the R\'enyi index and $k$ an additional replica index needed to address the randomness of the measurement outcomes. This method of analytically computing Born-weighted sums of nonlinear observables—such as entanglement entropy—is inspired from field of measurement-induced criticality~\cite{PhysRevB.100.134203, PhysRevB.101.104301, PhysRevB.101.104302,PhysRevB.99.174205}. In our case, $\mathcal{Z}_A $ and $\mathcal{Z}_0$  readily admit a Euclidean path integral representation as partition functions in replicated space. The ground-state density matrix is given by $\rho \propto \lim_{\beta \to \infty} e^{-\beta \hat{H}}$, with $\tr \rho = Z = \int \mathcal{D}[\varphi] e^{-S[\varphi]}\equiv 1$, where the path integral is subject to periodic boundary conditions in the imaginary time direction. Thus, the field $\varphi$ lives on an infinite cylinder with circumference $L$. Non-normalized post-measurement states $\rho_{\bf m}$ correspond to constrained path integrals in which the field is pinned to the measurement outcomes, leading to $\tr \rho_{\bf m} = Z_{\bf m} = \int \mathcal{D}[\varphi] e^{-S[\varphi]} \delta(\varphi(x,\tau=0) - \mathbf{m}(x))$, where $\mathbf{m}(x)$ denotes the continuum measurement result. Such a pinning of fields in the measurement region $C$ can be illustrated as defects/slits on the cylinder (see Figure~\ref{Fig1}). Accordingly, $\mathcal{Z}_0$ consists of $Q = nk + 1$ replicas of $\tr \rho_{\bf m}$, all constrained to the same measurement outcome $\mathbf{m}(x)$. In contrast, $\mathcal{Z}_A$ includes $k$ replicas of the form $\tr\rho^n_{{\bf m},A}$, along with a single “Born” replica of the form $\tr\rho_{\bf m}$. {To construct $\tr\rho^n_{{\bf m},A}$, one first traces out region $B$ by gluing the fields at $\tau=0^+$ and $\tau=0^-$ within $B$ (or equivalently, in the bra and ket of $\rho_{\bf m}$), and then cyclically glues $n$ such copies along region $A$, thereby implementing both the $n$-fold product and the final trace~\cite{Pasquale_Calabrese_2004, calabrese2009entanglement}. Graphically, this corresponds to the field living on an $n-$sheeted Riemann cylinder $\mc M_n$, with measurement-induced defects on each of the $n$ sheets and a branch cut along region $A$ resulting from the cyclic gluing. As a result, we can write $\tr \rho^n_{\bf{m}, A} = Z_{\mathbf{m}, \mc M_n}$, where  $Z_{\bf{m}, \mc M_n}$ is the $n-$sheeted Riemann surface version of $Z_{\bf{m}}$ with a branch cut in region A.
\begin{figure*}[t]
    \centering
    \includegraphics[width=\textwidth]{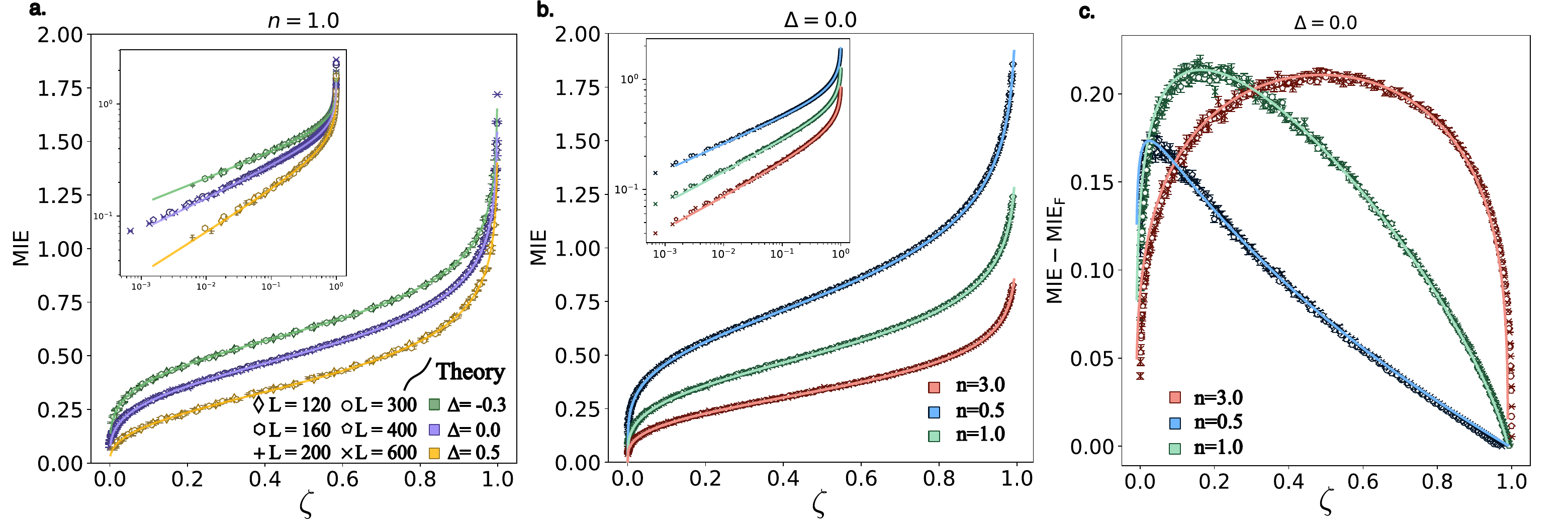}
    \caption{\textbf{MIE collapses vs cross ratio $\crat$ in the XXZ chain.} Markers denote numerical results, while solid lines represent theoretical predictions. Insets show the log-log version of the same plot where the low cross ratio $\zeta\ll 1$ regime is emphasized. \textbf{(a) }MIE for interaction strengths $\Delta = -0.3$, $0.0$, and $0.5$. \textbf{(b)} MIE for the XX chain ($\Delta = 0$) at different Rényi indices $n = 0.5$, $1.0$, and $3.0$. \textbf{(c)} Difference $\mathrm{MIE} - \mief$ for the XX chain for the same values of $n$.}
    \label{plots}
\end{figure*}

\textit{Conformal map and quantum-classical split---} 
Anticipating the conformal invariance of MIE observed numerically~\cite{Lin2023probingsign,PhysRevB.109.195128}, we first conformally map $\mathcal M_n$ onto a finite cylinder $\mc C(n)$ of height $h(\crat)/n$ and circumference $\beta = 2\pi$, where $\crat$ is the cross-ratio (Fig.~\ref{Fig1})~\cite{kythe2019handbook,Rajabpour_2016,antonini_holographic_2022}. The height obeys $h(\crat)=2\pi\mathcal K(k)/\mathcal K(\sqrt{1-k^{2}})$, with $k=(1-\sqrt{1-\crat})/(1+\sqrt{1-\crat})$ and $\mathcal K$ the complete elliptic integral of the first kind. This mapping introduces a geometric free energy contribution $F^{\mathrm{geom}}_n=-\log Z^{\mathrm{geom}}_{n}$, which factors out of $Z_{\mathbf m,\mathcal M_n}$ \cite{Bimonte_2013,Rajabpour_2016}. Next, we decompose the bosonic field $\varphi=\varphi_q+\varphi_{\mathrm{cl},\mathbf m}$ into a classical solution $\varphi_{\mathrm{cl},\mathbf m}$ that satisfies the boundary conditions encoding the measurement outcomes (up to winding), and ``quantum fluctuations'' satisfying Dirichlet boundary conditions. Due to the Gaussianity of the free-boson action, a corresponding split occurs at the level of the partition functions too, giving 
\begin{equation}
    Z_{\bf{m},\mc M_n}=\frac{Z^{\rm geom}_{n} }{\eta(q_n)} \sum_w \exp[-S_{\mc C(n)}[\vphi_{\rm cl,\bf m}]],
\end{equation}
where $q_n = e^{-\pi \beta/(h/n)} = e^{-2\pi^2 n/h}$,  
$\eta(q)= q^{1/24} \prod_{s=1}^\infty (1 -q^s)$ is the Dedekind eta function capturing the ``quantum'' contribution $Z_{D, \mc C(n)}^q =1/\eta(q_n)$~\cite{DiFrancesco1997}, while $\exp[-S_{\mc C(n)}(\varphi_{\mathrm{cl},\mathbf m})]$ captures the classical contribution on $\mc C(n)$ with $\sum_w$ representing the sum over topological sectors. The $\bf m$-independent terms above lead to a trivial simplification of the replica limit since the replicas decouple~\cite{supplement}. Furthermore, we find that the geometric term vanishes in \eqref{replicalimit}, i.e., $\log[Z^{\mathrm{geom}}_{n}/(Z^{\mathrm{geom}}_{1})^{n}]=0$ due to the Rényi structure of the problem~\cite{supplement}, leaving behind only the quantum and classical contributions.

\textit{Winding contributions and the MIE --- } The quantum contribution appears equally to MIE and $\mief$ since it is independent of $\bf m$; the distinction therefore arises solely from the classical part that carries measurement dependence. In the forced case where we post-select to $\ket{01010\dots}$, the measurement outcomes flow to the same Dirichlet boundary conditions on both boundaries, leading to the classical part $ \sum_w \exp[-S_{\mc C(n)}[\vphi_{{\rm cl},D}]] = \sum_w q_n^{gw^2}$, which is simply the classical part in the cylinder partition function $Z_{D,\mc C(n)}$. 
For MIE on the other hand, the replica trick~\eqref{replicalimit} couples different replicas in both $\mc Z_A$ and $\mc Z_0$ leading to a non-trivial contribution. For example, the $Q$ replicas in $\mc Z_0$ are coupled via matching boundary conditions on the measured region up to winding, {\it i.e. } $\vphi^{(i)}_{{\rm cl},\bf m}\vert_{C}= \mathbf{m}+ 2\pi w_i,\quad i = 1,\dots,Q$, where $w_i\in\mathbb{Z}$ are the winding numbers for each replica. Such a coupling is handled by a trick introduced by Fradkin and Moore~\cite{PhysRevLett.97.050404} in a different context: since the fields $\vphi^{(i)}_{{\rm cl},\mathbf{m}}$ take the same values in region $C$ (modulo winding), one can rotate to a new basis $\bar{\vphi}^{(i)}_{{\rm cl}}$ such that $Q-1$ of them have boundary conditions that vanish (modulo compactification) in $C$,  while only a single ``center-of-mass'' (c.o.m) field $\bar{\vphi}^{(Q)}_{{\rm cl},\mathbf{m}} =\tfrac{1}{\sqrt{Q}} \sum_i \vphi^i_{{\rm cl},\bf m}$ retains the explicit dependence on $\mathbf{m}$. Then, the first $Q-1$ replicas contribute through a summation over different winding sectors $\sum_{\{w_i\}\in \mathbb{Z}^{Q-1}}\exp[-\sum_{i=1}^{Q-1}S[\bar{\vphi}^{i}_{{\rm cl}}]]$ in this rotated basis, while the last measurement-dependent c.o.m replica is trivial to evaluate and vanishes in the replica limit. Such winding sums—neglected in the original Fradkin-Moore treatment~\cite{PhysRevLett.97.050404}, but later reinstated in several works~\cite{PhysRevB.79.115421,oshikawa2010boundary, PhysRevLett.107.020402, Zhou_2016}—are essential for properly accounting for compactification in the new basis and are thus key to correctly evaluating $\mc Z_0$ and $\mc Z_A$, and hence, the MIE. The calculation for $\mc Z_A$ proceeds similarly but requires a generalization of the above trick. We defer the detailed derivation of these to the supplementary material~\cite{supplement} and directly give the winding contribution for $\mc Z_A$: 
\begin{equation}
\label{windingint}
\mc W_{n,k}(\crat,g) = \sqrt{\frac{Qg}{2\pi h}}\int_{-\infty}^{\infty}d\delta_\vphi \;e^{- \frac{g\delta_{\vphi}^2}{2h}}\left[\sum_{w\in\mathbb{Z}}q_n^{g\left(w+\frac{\delta_\vphi}{2\pi}\right)^2}\right]^k,
\end{equation}
where we have emphasized its dependence on $\crat$ and $g$. The above can be used to calculate the MIE, which we derive to be~\cite{supplement}
\begin{equation}
\label{MIEnmainresult}
    \mathrm{MIE}^{(n)}(A) =
\frac{1}{1-n} \left[\mc W'_n -n\mc W'_1-\log\frac{\eta(q_n)}{\eta(q_1)^n}  \right],
\end{equation}
where $\mathcal{W}'_n =\lim_{k \ra 0}\partial_k \mathcal{W}_{n,k}$. The analyticity of $\mc W_{n,k}$ allows one to take both the $k\ra0$ and $n\ra 1$
limit and evaluate MIE for all $n$ exactly. 
The winding contribution~\eqref{windingint} has an appealing interpretation as an average over all possible Dirichlet boundary conditions $(\vphi_1,\vphi_2)$ at $(C_1,C_2)$ indexed by $\delta_{\vphi}=\vphi_1-\vphi_2 \in [0, 2\pi)$, with weight given by the corresponding partition function $Z_{\mc C(1),\delta_\varphi} \sim  \sum_{w\in\mathbb{Z}}q_1^{g\left(w+\delta_\vphi/(2\pi)\right)^2}$, as expected from ``Born averaging''~\cite{supplement}. The emergence of Dirichlet boundary conditions is natural given that these boundary conditions form fixed points of the free-boson BCFT and respect the symmetries preserved by the measurements.   
The winding contribution for the forced case can be viewed as the special case $\delta_\vphi=0$ in \eqref{windingint}. Both $\mathrm{MIE}$ and $\mathrm{MIE}_{\rm F}$ are universal and conformally-invariant, with all cross‑ratio dependence in~\eqref{MIEnmainresult} entering solely through the function $h(\crat)$, despite the existence of specific measurement outcomes that are known to break conformal invariance~\cite{Stéphan_2014,PhysRevB.93.125139}.

\textit{Asymptotics ---} Since MIE quantifies the entanglement between two distant regions after the rest of the system has been measured, the regime of primary interest is $\crat \ll 1$, i.e., when the unmeasured regions $A$ and $B$ are maximally separated. In this regime, the leading contribution to the MIE comes from the new winding terms we calculated. Although the derivative of the integral~\eqref{windingint} cannot be evaluated analytically, its asymptotic behavior can still be extracted. We find that the MIE exhibits two notable features. First, it undergoes a qualitative change in behavior at $n = 1/2$. For $n < 1/2$, $\mathrm{MIE} \sim \crat^{2n(1-n)g}$, while for $n \geq 1/2$, the exponent saturates to $\crat^{g/2}$ with an $n$-dependent pre-factor. This contrasts with the forced MIE, which scales as $\crat^{2ng}$ for $n < 1$ and saturates to $\crat^{2g}$ for $n > 1$~\cite{Rajabpour_2016}. Second, for $n > 1/2$, the leading term includes a $1/\sqrt{\log(1/\crat)}$ pre-factor, arising from the replica limit $k \to 0$. While logarithmic terms are common in logarithmic CFTs~\cite{GURARIE1993535, doi:10.1142/9789812775344_0032, vasseur2012logarithmic, cardy1999logarithmic}, and integer-powered logarithms appear in certain entanglement entropy expansions at small $\crat$~\cite{rajabpour2012entanglement, ruggiero2018entanglement}, a $1/\sqrt{\log (1/\crat)}$ factor is unusual and does not arise from standard operator product expansions. Finally, we note that MIE includes contributions from pre-existing entanglement. Ref.~\cite{PhysRevB.109.195128} defines the measurement-induced information (MII), a related quantity which subtracts the pre-measurement mutual information from the measurement-averaged value. The scalings derived here show that the MII is positive for \textit{real} measurements and becomes negative in the forced case, further highlighting the inadequacy of the latter as a proxy for the former~\cite{supplement}.

\textit{Numerical Results --- } We now present numerical evidence supporting our claims. As a model, we consider the periodic XXZ spin-1/2 chain with Hamiltonian
\begin{equation}
    H = \sum_j \sigma^x_j\sigma^x_{j+1} + \sigma^y_j\sigma^y_{j+1} + \Delta\sigma^z_j\sigma^z_{j+1}, 
\end{equation}
where $\Delta$ tunes the interaction strength. This model hosts a gapless phase for $\Delta \in (-1,1]$, with low-energy physics described by a TLL, where $\Delta = -\cos(\pi g)$ sets the Luttinger parameter~\cite{giamarchi_quantum_2003}. At $\Delta = 0$, the model maps to free fermions, allowing for exact entanglement entropy calculations~\cite{Peschel_2009}. For $\Delta \neq 0$, we use the \texttt{iTensor} library \cite{itensor} to obtain approximate ground states using the DMRG algorithm~\cite{white1992density, white1993density}, from which entanglement entropy is readily computed. To avoid parity effects, we use a symmetric setup: both measured regions are of equal length and placed antipodally. We sample measurement outcomes in the $\sigma_z$ basis according to the Born rule, compute the entanglement of region $A$, and average over trajectories to obtain the MIE for a given $\crat$. By increasing the length of the measured regions while preserving their symmetric placement, we access a range of cross-ratios $\crat \in (0,1)$. For the forced case, we post-select to the antiferromagnetic state $\ket{\uparrow\downarrow\uparrow\cdots}$, which is known to flow to a Dirichlet boundary condition~\cite{giamarchi_quantum_2003}. Our numerical results are shown in Fig.~\ref{plots}.
As seen in the plots, the MIE shows clear dependence on both $n$ and $\Delta$, in excellent agreement with theoretical predictions. 

\textit{Summary and Discussion --- } We have presented an exact calculation of the MIE for a broad class of quantum critical states in 1+1D—namely, Luttinger liquids described at low energies by a compactified free boson CFT. We find that the MIE is a universal function of the cross-ratio $\crat$. To obtain this, we used a replica trick approach to handle the intrinsic randomness introduced by measurements and showed that the leading universal contributions arise from carefully tracking winding numbers across replicas. While winding sectors are common in compact boson calculations, their role here specifically dictated by the nature of the measured operator (charge) and the Born-rule weighting of outcomes. Our results thus highlight that \textit{physical} measurements differ qualitatively from post-selection on fixed outcomes—a proxy often used in earlier works~\cite{PhysRevB.92.075108, Rajabpour_2016, najafi_entanglement_2016, PhysRevB.111.155143}. Crucially, our final result for the MIE has a natural interpretation in terms of summing over Dirichlet boundary conditions at $C_1$ and $C_2$, weighted by the corresponding partition function, consistent with Born averaging. We further support our theoretical predictions with numerical simulations on the XXZ spin chain, finding excellent agreement. 

A natural continuation of our work would be to compute MIE in other CFTs and measuring other operators to check whether our interepretation in terms of Born averaging over conformal boundary conditions holds. In particular, we anticipate that the MIE could provide a valuable tool to diagnose symmetry-enriched CFTs and gapless symmetry protected topological phases~\cite{PhysRevX.7.041048,PhysRevX.11.041059, PhysRevB.104.075132}, see~\cite{Lin2023probingsign}. While we focused in this letter on the MIE averaged over quantum trajectories~\eqref{miedef}, our approach can be generalized to compute higher cumulants and the full distribution of MIE over measurement outcomes~\cite{InPrep}.
Finally, it would also be interesting to study whether the MIE has a natural holographic description (see Refs.~\onlinecite{antonini_holographic_2022, antonini_holographic_2023, brinster_measurement_2025} for forced measurements), which could be generalized to higher dimensions. 

{\it Data availability --} The data and codes used to generate Fig.~\ref{plots} and Fig.~1 in the supplementary material \cite{supplement} are available on
Zenodo~\cite{khanna_2025_19359106}.

 {\it Acknowledgments -- } This work was supported in part by the US Department of Energy, Office of Science, Basic Energy Sciences, under award No. DE-SC0023999, and by the Swiss National Science Foundation (grant 10008234). 
 We thank Zihan Cheng, Sarang Gopalakrishnan and Andrew Potter for useful discussions at earlier stages of this project, and Sara Murciano for valuable comments on this manuscript.

% \bibliographystyle{unsrt}

% \bibliography{apssamp}% Produces the bibliography via BibTeX.
% \input{main.bbl}
% % Adding sup mat
\bigskip
\onecolumngrid
\newpage
\includepdf[pages=1]{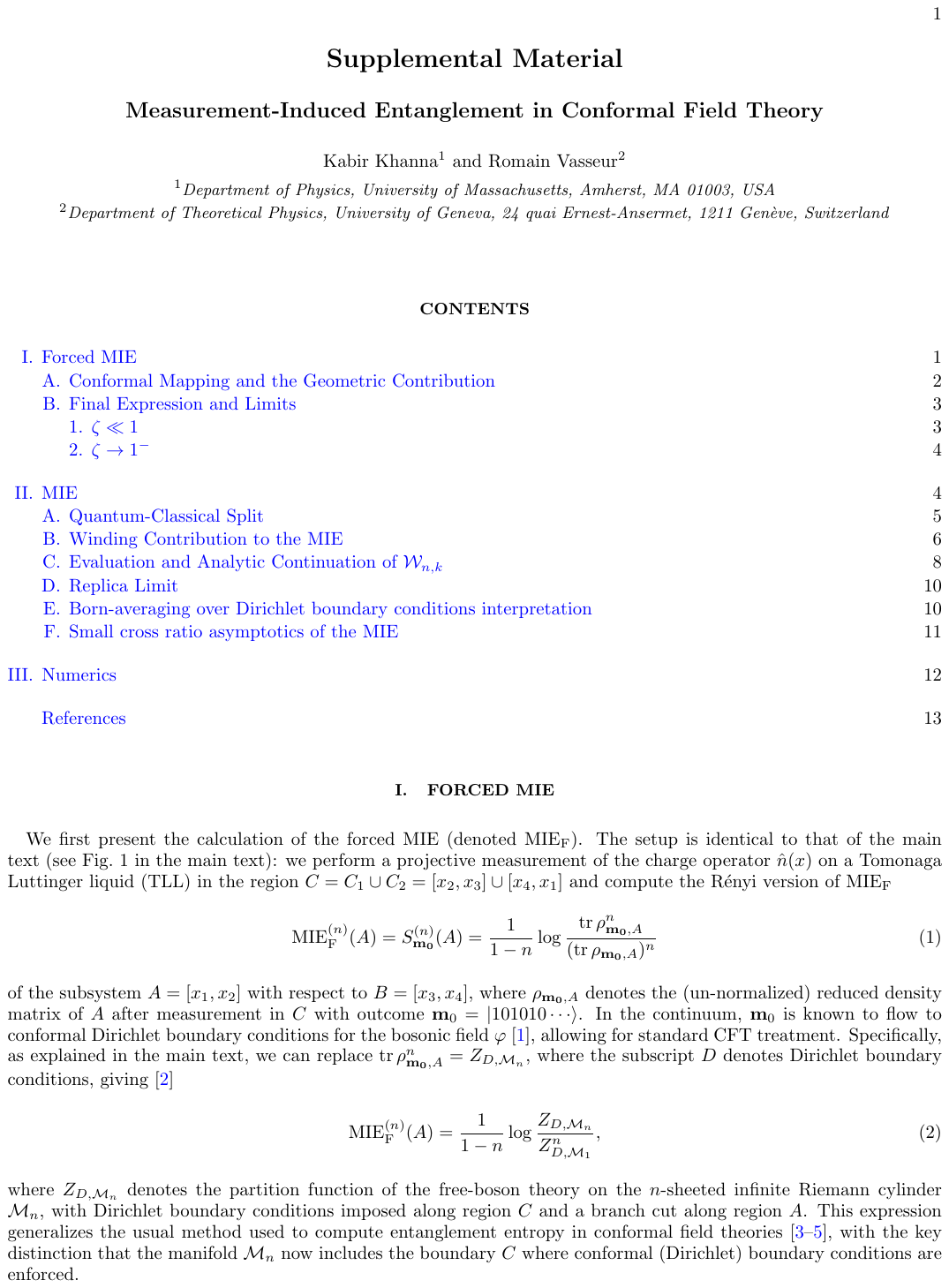}
\newpage
\includepdf[pages=2]{suppmat.pdf}
\newpage
\includepdf[pages=3]{suppmat.pdf}
\newpage
\includepdf[pages=4]{suppmat.pdf}
\newpage
\includepdf[pages=5]{suppmat.pdf}
\newpage
\includepdf[pages=6]{suppmat.pdf}
\newpage
\includepdf[pages=7]{suppmat.pdf}
\newpage
\includepdf[pages=8]{suppmat.pdf}
\newpage
\includepdf[pages=9]{suppmat.pdf}
\newpage
\includepdf[pages=10]{suppmat.pdf}
\newpage
\includepdf[pages=11]{suppmat.pdf}
\newpage
\includepdf[pages=12]{suppmat.pdf}
\newpage
\includepdf[pages=13]{suppmat.pdf}
\newpage
\includepdf[pages=14]{suppmat.pdf}

\end{document}

\begin{center}
\textbf{\Large Supplemental Material\\
\vspace{1em}
\large Measurement-Induced Entanglement in Conformal Field Theory}

\vspace{5mm}

Kabir Khanna\textsuperscript{1} and Romain Vasseur\textsuperscript{2}
% \author{Kabir Khanna}
% \affiliation{Department of Physics, University of Massachusetts, Amherst, Massachusetts 01003, USA}

% \author{Romain Vasseur}%
%  %\email{romain.vasseur@unige.ch}
%  \affiliation{Department of Theoretical Physics, University of Geneva, 24 quai Ernest-Ansermet, 1211 Gen\`eve, Switzerland}
\vspace{1mm}

\textsuperscript{1}\textit{\small Department of Physics, University of Massachusetts, Amherst, MA 01003, USA}

\textsuperscript{2}\textit{\small Department of Theoretical Physics, University of Geneva, 24 quai Ernest-Ansermet, 1211 Gen\`eve, Switzerland}

\end{center}

\makeatother

\tableofcontents

\section{Forced MIE}
\label{forcedmiesection}
We first present the calculation of the forced MIE (denoted $\mief$). The setup is identical to that of the main text (see Fig.~1 in the main text): we perform a projective measurement of the charge operator $\hat{n}(x)$ on a Tomonaga Luttinger liquid (TLL) in the region $C = C_1\cup C_2 =  [x_2, x_3] \cup [x_4, x_1]$ and compute the Rényi version of $\mief$
\begin{equation}
\miefn(A) = S^{(n)}_{\bf m_0}(A) = \frac{1}{1-n}\log\frac{\tr \rho^n_{\mathbf{m_0}, A}}{(\tr\rho_{\mathbf{m_0}, A})^n}
\end{equation}
of the subsystem $A = [x_1, x_2]$ with respect to $B = [x_3, x_4]$, where $\rho_{\mathbf{m_0}, A}$ denotes the (un-normalized) reduced density matrix of $A$ after measurement in $C$ with outcome $\mathbf{m}_0=\ket{101010\cdots}$. In the continuum, $\mathbf{m}_0$ is known to flow to conformal Dirichlet boundary conditions for the bosonic field $\vphi$~\cite{giamarchi_quantum_2003}, allowing for standard CFT treatment. Specifically, as explained in the main text, we can replace $\tr \rho^n_{\mathbf{m_0}, A} = Z_{D,\mc M_n}$, where the subscript $D$ denotes Dirichlet boundary conditions, giving \cite{Rajabpour_2016}
\begin{equation}
\label{miefpartitionratio}
\miefn(A) = \frac{1}{1-n} \log\frac{Z_{D,\mc M_n}}{Z_{D,\mc M_1}^n},
\end{equation}
where $Z_{D,\mc M_n}$ denotes the partition function of the free-boson theory on the $n$-sheeted infinite Riemann cylinder $\mc M_n$, with Dirichlet boundary conditions imposed along region $C$ and a branch cut along region $A$. This expression generalizes the usual method used to compute entanglement entropy in conformal field theories \cite{HOLZHEY1994443, Pasquale_Calabrese_2004, calabrese2009entanglement}, with the key distinction that the manifold $\mc M_n$ now includes the boundary $C$ where conformal (Dirichlet) boundary conditions are enforced. 

\subsection{Conformal Mapping and the Geometric Contribution}
\label{conformalmappingforcedmie}
Just like in the main text, the first step in simplifying $Z_{D,\mc M_n}$ is to conformally map the Riemann surface $\mc M_n$ on the $z = x+i\tau$ plane to a finite cylinder $\mc C(n)$ through a map $\bar{w}_n(z)$~\cite{kythe2019handbook,Rajabpour_2016,antonini_holographic_2022} (see Fig.~1 in the main text). Such a map is constructed as follows: first, we map the $n-$sheeted Riemann cylinder with slits (measured regions) to the $n-$sheeted Riemann plane with slits symmetrically placed about the imaginary time axis through a map $\tilde z(z)$. Next, we map this to the $n-$sheeted circular strip (annulus) via a map $w(\tilde z)$ and uniformize through $w_n(\tilde z) = w(\tilde{z})^{1/n}$. Finally, we map from the annulus to the cylinder $\mc C(n)$ with $\bar{w}_n(w_n) = \log w_n$. The resulting cylinder has a circumference $\beta = 2\pi $ and height $h_n(\crat) =h(\crat)/n$ where
\begin{equation}
\label{hxdef}
    h(\crat) = 2\pi \frac{\mc K (k)}{\mc K(\sqrt{1-k^2})}. 
\end{equation}
Here, $k = (1 - \sqrt{1 - \crat})(1 + \sqrt{1 - \crat})^{-1}$, with $\crat = w_{12}w_{34}w^{-1}_{13} w^{-1}_{24}$ being the cross-ratio, and $w_{ij} = (L/\pi) \sin\left(\pi x_{ij}/L \right)$ denotes the chord length for a system of size $L$. The function $\mathcal{K}(k)$ is the complete elliptic integral of the first kind and is defined as
\begin{equation}
    \mc K(k) = \int_0^{\pi/2} \frac{d \theta}{\sqrt{1-k^2\sin^2\theta}}.
\end{equation}
Note that the overall Rényi dependence after mapping to the cylinder is $\bar{w}_n(z) = (1/n)\log(w(\tilde z(z)))$. Since our analysis only requires this extracted Rényi dependence, we omit the explicit forms of the intermediate maps $w(\tilde{z})$ and $\tilde{z}(z)$ (for a detailed construction, the reader may refer to Refs.~\cite{Rajabpour_2016, antonini_holographic_2022}). This conformal mapping induces a change in the free energy $\delta F_{D,\mc M_n} = -\delta \log Z_{D, \mc M_n}$, given by \cite{DiFrancesco1997}
\begin{equation}
\label{deltaf}
    \delta F_{D, \mc M_n} = \frac{i\delta l}{2\pi} \oint_{C_2}\langle T(z)\rangle dz \; +\; \text{c.c.}, 
\end{equation}
where $T(z)$ is the energy-momentum tensor on $\mc M_n$, $\delta l$ is a small deformation of the interval $A$, and the contour encircles the region $C_2$ counter-clockwise. To simplify this expression, we use the transformation law of the stress tensor under conformal maps: $T(z) = (\partial_z \bar{w}_n)^2 T(\bar{w}_n)+(c/12)\{\bar{w}_n,z\}$, where $c$ is the central charge and $\{f,z\} = (f'''/f') - (3/2)(f''/f')^2$ is the Schwarzian derivative. On the cylinder with coordinate $\bar{w}_n$, the variation of the free energy with respect to the cylinder height $h_n$ satisfies $\delta F_{\mc C(n)}/\delta h_n = 2\langle T(\bar{w}_n)\rangle$ and the height deformation under slit displacement is given by $\delta h_n/\delta l = (i/2\pi)\oint_{C_2}(\partial_z \bar{w}_n(z))^2$ \cite{DiFrancesco1997, Bimonte_2013, Rajabpour_2016}. Substituting these into \eqref{deltaf}, we find that the free energy on $\mathcal{M}_n$ splits into two contributions \cite{Bimonte_2013, Rajabpour_2016}:
\begin{equation}
\label{freeenergysplitforcedmie}
    F_{D, \mc M_n} = F_{D, \mc C(n)} + F^{\mathrm{geom}}_n, 
\end{equation}
where $F^{\mathrm{geom}}_n=-\log Z^{\mathrm{geom}}_n$ is the contribution purely due to the geometry of the map $\bar{w}_n$ and the central charge and $F_{D, \mc C(n)}$ is the free energy on the cylinder $\mc C(n)$ with Dirichlet boundary conditions on both boundaries. The former satisfies
\begin{equation}
\label{geometricfreeenergydiff}
    \frac{\delta F^{\mathrm{geom}}_n}{\delta l} = \frac{i}{12\pi} \oint_{C_2} \{\bar{w}_n,z\}\; dz.
\end{equation}
Re-writing \eqref{miefpartitionratio} in terms of these free energies, we see that $\mief$ also splits into two parts,
\begin{equation}
\label{freemiefgeocyl}
    \miefn(A) = \frac{1}{n-1}[F^{\mathrm{geom}}_n - nF^{\mathrm{geom}}_1] +\frac{1}{n-1}[F_{D, \mc C(n)} - nF_{D, \mc C(1)}], 
\end{equation}
where the former encodes the geometric contribution and latter encodes the universal operator content of the compactified free-boson theory on the cylinder $\mc C(n)$. Next, since $\bar{w}_n(z)=(1/n)\log(w(\tilde z(z)))$ the Rényi dependence drops out entirely when evaluating the Schwarzian derivative $\{\bar{w}_n, z\}$. This follows from the invariance of the Schwarzian under rescaling, i.e., $\{ \lambda \bar{w_n}, z \} = \{ \bar{w_n}, z \}$ for any constant $\lambda$. Additionally, because the contour integral is evaluated via the summation of residues at the poles of the Schwarzian located at $x_1$ and $x_4$, and these poles are repeated $n$ times on the $n$-sheeted surface $\mathcal{M}_n$, we obtain an overall factor of $n$, i.e., $F^{\mathrm{geom}}_n = n F^{\mathrm{geom}}_1$. This implies that the geometric contribution to $\mief$ vanishes in \eqref{freemiefgeocyl}, leading to:
\begin{equation}
\label{mieffinal}
    \miefn(A) = \frac{1}{n-1}[F_{D,\mc C(n)} - nF_{D,\mc C(1)}] = \frac{1}{1-n}\log{\frac{Z_{D,\mc C(n)}}{Z_{D,\mc C(1)}^n}}, 
\end{equation}
where $Z_{D,\mc C(n)}=e^{-F_{D, \mc C(n)}}$ is the partition function on the cylinder $\mc C(n)$ with the \textit{same} Dirichlet boundary conditions on both the boundaries. We note that the vanishing of the geometric contribution arises specifically from the Rényi structure of our problem. It does not occur, for instance, in the original context where this term was introduced—namely, in the computation of Casimir interaction between surfaces in a medium \cite{Bimonte_2013}—where such a structure is absent. A few existing works \cite{Rajabpour_2016, najafi_entanglement_2016, antonini_holographic_2023} that encounter the geometric contribution while calculating entanglement measures such as $\mief$ do not emphasize this and proceed to explicitly evaluate the geometric term, despite it ultimately vanishing when considered in the manner above. Unlike here, these works evaluate free energies on the annulus instead of the cylinder, making their geometric contribution differ from ours by a universal constant term. In particular, $F^{\mathrm{geom}}_{\mc A(n)} = F^{\mathrm{geom}}_n - \frac{ch}{12n}$, where $F^{\mathrm{geom}}_{\mc A(n)}$ is the analog of $F^{\mathrm{geom}}_n$ but evaluated on the annulus with radii $e^{-h/n}$ and $1$. The claim we make here is that $F^{\mathrm{geom}}_n$ does not contribute to $\mathrm{MIE_F}$. We note that this is not in contradiction with Refs.~\cite{Rajabpour_2016, najafi_entanglement_2016, antonini_holographic_2023}, since any leading contribution attributed to the “geometric” term in those works indeed comes from the universal $\frac{ch}{12n}$ piece in $F^{\mathrm{geom}}_{\mc A(n)}$. This in turn precludes Refs.~\cite{Rajabpour_2016, najafi_entanglement_2016, antonini_holographic_2023}
 from writing a closed form expression for the $\mief$, restricting those works to perturbative expansions in special limits of the $\mief$. Our analysis clarifies that the geometric part for the $\mief$ vanishes entirely, thereby simplifying the the computation and giving a closed expression (that we write in a moment), which depends only on the ratio of standard partition functions on the cylinder—making its universal and conformally-invariant nature manifest. The nontrivial contribution that does however enter due to the conformal mappings is the height of the cylinder~\eqref{hxdef} (up to a $1/n$ factor). This quantity depends non-trivially (but only) on the cross-ratio $\crat$ which proves the conformal invariance of $\mief$.

\subsection{Final Expression and Limits}
\label{finalexpforcedmie}
Using the explicit form~\cite{DiFrancesco1997} of the compact free boson partition functions in \eqref{mieffinal}, we get
\begin{equation}
\label{finalmief}
    \miefn(A)  = \frac{1}{1-n} \left[\log\frac{{\sum_{w\in\mathbb{Z}}q_n^{gw^2}}}{\left(\sum_{w\in\mathbb{Z}}q_1^{gw^2}\right)^n}-\log\frac{\eta(q_n)}{\eta(q_1)^n}\right],
\end{equation}
Here $q_n = e^{-\pi \beta/(h(\crat)/n)} = e^{-2\pi^2n/h(\crat)}$, and  $\eta(q) = q^{1/24}\prod_{s=1}^{\infty}(1-q^s)$ is the Dedekind eta function. In the language of the main text, the second term above which contains a ratio of Dedekind eta functions, with 
\begin{equation}
Z^q_{D, \mc C(n)} = {\rm e}^{-\beta E_0} \prod_{s=1}^\infty \frac{1}{1-{\rm e}^{- \beta \omega_s}} = \frac{1}{\eta(q_n)},
\end{equation}
captures the ``quantum fluctuations" contribution to the full partition function $Z_{D, \mc C(n)}$, with $\beta=2\pi$, $\omega_s = s \pi/(h/n)$ the energy of the free boson modes on the cylinder, and $E_0 = \frac{1}{2} \sum_{s=1}^{\infty} \omega_s = - \frac{\pi}{24(h/n)}$ the ground-state (Casimir) energy. On the other hand, the term with the summations $\sum_{w\in\mathbb Z}$ can be interpreted as summing over the winding sectors of the (exponentiated) action of the classical solutions $ \vphi_{\rm cl,\m_0} (x,\tau) = \m_0 + 2 \pi w x / (h/n)$ with the same Dirichlet boundary conditions ($\m_0$, which is uniform and fixed by the forced measurement outcomes) on both the boundaries of the cylinder, making this the ``classical" winding contribution for $\mief$ as described in the main text:
\begin{equation}
\label{classicalpartforced}
\sum_w \exp[-S_{\mc C(n)}[\vphi_{\rm cl,\bf m_0}]] = \sum_{w \in {\mathbb{Z}}} \exp \left[- \frac{g}{4 \pi} \int_0^{\beta = 2 \pi} d \tau \int_0^{h/n} dy \left(\frac{2 \pi w }{ (h/n)}\right)^2  \right] = \sum_{w \in {\mathbb{Z}}} q_n^{g w^2}.
\end{equation}
Next, while the various limits of $\mief$ have already been discussed in Ref.~\cite{Rajabpour_2016}, we reproduce them here for completeness, as our notation and computations differ from those used in that work.
\subsubsection{$\crat\ll 1$}
\label{forcedmiexll1}
This is the limit in which the measurements create two distant regions, making $\mief$ a useful diagnostic to study. In this limit, $h(\crat) = \pi^2/\log(16/\crat) + {\cal O}(\crat/\log^2(\crat))$, $q_n = (\crat/16)^{2n}$, and the leading term to the $\miefn$ comes from the $w=1$ sector of the classical contribution, simplifying \eqref{finalmief} (to this leading order) as: 
\begin{equation}
    \mief^{(n)}(A) = \frac{1}{1-n}\left[\left(\frac{\crat}{16}\right)^{2ng} - n\left(\frac{\crat}{16}\right)^{2g} \right]+\dots
\end{equation}
resulting in the scaling at small $\crat$
\begin{equation}
\miefn \underset{\crat \to 0}{\sim}
\begin{cases}

\displaystyle 
\crat^{2g}
& n > 1,\\[1ex]
\displaystyle
\crat^{2g}\log \crat
& n = 1, \\[1ex]
\displaystyle
\crat^{2ng}
& n< 1, 
\end{cases}
\end{equation}
where we dropped unimportant prefactors. 

\subsubsection{$\crat \to 1^{-}$}
\label{x1forcedmie}
This is the limit of small measurement regions. Intuitively, measuring very few sites should return the usual scaling of entanglement entropy in a conformal field theory which reads $S(A) = \tfrac{c(n+1)}{6n}\log ((L/\pi a)\sin(\pi l/L))$, where $c$ is the central charge, $l$ is the length of the interval $A$, and $a$ is the UV cutoff \cite{HOLZHEY1994443, Pasquale_Calabrese_2004, calabrese2009entanglement}, making this limit a fundamental benchmark of our results. In this limit, it is best to first perform a modular transformation on $Z_{D, \mc C(n)}$ under which it is invariant \cite{DiFrancesco1997}, giving us
\begin{equation}
\label{modulartrasformedpartition}
     Z_{D, \mc C(n)} = \frac{1}{\sqrt{2g}}\frac{1}{\eta(\tilde q_n)}\sum_{w\in\mathbb{Z}} (\tilde q_n)^{w^2/4g},
\end{equation}
where $\tilde q_n = e^{-\frac{4\pi}{\beta}\frac{h(\crat)}{n}} = e^{-2h(\crat)/n}$, and $h(\crat) = \log(16/(1-\crat))$ for $\crat\ra 1^-$. The summation over the winding sectors simplifies to a power series in $(1-\crat)$ when substituting \eqref{modulartrasformedpartition} in \eqref{mieffinal}  while the Dedekind eta function gives the leading logarithmic contribution (through the $\tilde q^{1/24}$ Casimir term), resulting in 
\begin{equation}
    \mief \underset{\crat \to 1^-}{\sim} \frac{1}{12}\frac{n+1}{n}\log\frac{1}{1-\crat}+\dots
\end{equation}
If we assume a symmetric setup with sizes of the regions $|A| =l, |B|=l,$ and $|C_1|=|C_2|=s$ with $s\ll l$, then $1/(1-\crat)= (L/\pi s)^2\sin^2(\pi l/L)$ and we get the expected entanglement scaling where $s$ plays the role of the UV cutoff.

\begin{figure}[t!]
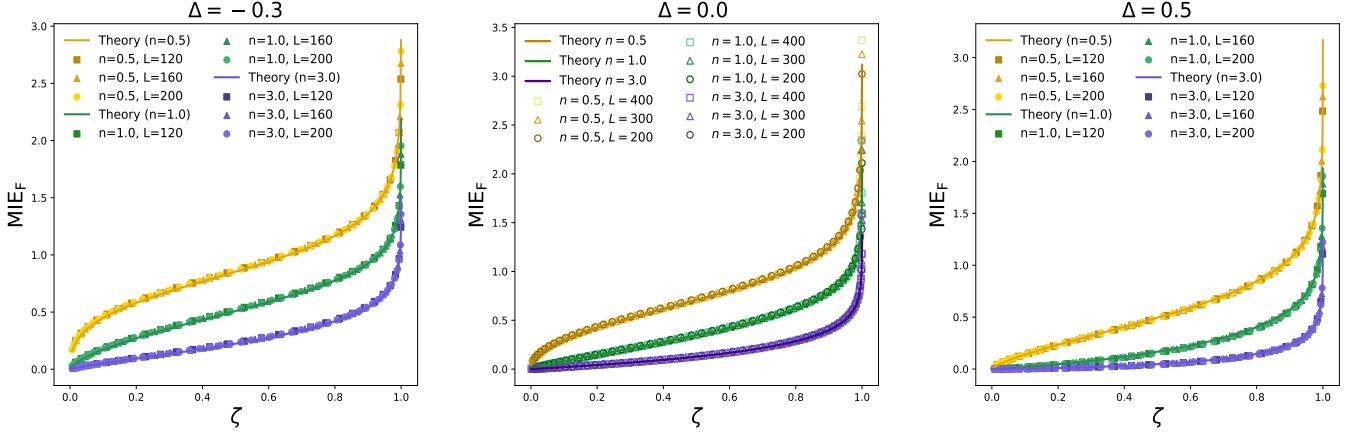

    \centering

    \begin{subfigure}[t]{0.32\textwidth}
        \centering
        \includegraphics[width=\linewidth, height=6cm]{Images/mieforced_Delta_-0.3_allrenyi.pdf}
        % \caption{Caption 1}
        \label{fig:subim1}
    \end{subfigure}%
    \hfill
    \begin{subfigure}[t]{0.32\textwidth}
        \centering
        \includegraphics[width=\linewidth, height=6cm]{Images/mieforced_Delta_0.0_allrenyi.pdf}
        % \caption{Caption 2}
        \label{fig:subim2}
    \end{subfigure}%
    \hfill
    \begin{subfigure}[t]{0.32\textwidth}
        \centering
        \includegraphics[width=\linewidth, height=6cm]{Images/mieforced_Delta_0.5_allrenyi.pdf}
        % \caption{Caption 3}
        \label{fig:subim3}
    \end{subfigure}
    \caption{\textbf{$\mathbf{MIE_F}$ collapses vs cross ratio $\mathbf{\zeta}$ in the XXZ chain.} Markers denote numerical results, while solid lines represent theoretical predictions. $\mief$ for interaction strength \textbf{(a) } $\Delta = -0.3$,  \textbf{(b) }$\Delta = 0$ and \textbf{(c) }$\Delta=0.5$, for Rényi indices $n=0.5,1.0$ and $3.0$.}
    \label{mieforcedfig}
\end{figure}
\section{MIE}
We now turn to a detailed derivation of the MIE—the central quantity in this work. We assume the same setup as in the forced case (see section \ref{forcedmiesection}) where instead of post-selecting to a specific outcome when measuring charge, we now average over all possible measurement outcomes and compute the Rényi version of the MIE as: 
\begin{equation}
\label{defmie}
    \mathrm{MIE}^{(n)}(A) = \sum_\mathbf{m}p_{\bf m} S^{(n)}_\mathbf m (A) =\sum_{\mathbf m} p_{\bf m}\frac{1}{1-n} \log\left[\frac{\tr \rho^{n}_{\mathbf{m}, A}}{(\tr\rho_{\mathbf m})^n}\right], 
\end{equation}
where $p_{\mathbf{m}}=\tr\rho_{\mathbf m}$ is the Born probability of outcome $\mathbf{m}$ and $\rho_{\mathbf{m}, A}$ denotes the (un-normalized) reduced density matrix of $A$ after measurement in $C$ with outcome $\mathbf{m}$. Unlike the forced MIE, the averaging over all measurement outcomes \textit{a priori} precludes us from  treating the boundary as a conformal boundary condition and writing an analog of \eqref{miefpartitionratio} for the MIE. However, we can use the replica trick
\begin{equation}
    \log x = \lim_{k\ra0}\frac{x^k-1}{k} = \lim_{k\ra 0}\frac{d}{dk}x^k,
\end{equation}
to re-write \eqref{defmie} in terms of effective replica partition functions as follows: 
\begin{align}
    \sum_\m p_\m \frac{1}{1-n}\log\bigg[\frac{\tr\rho^n_{\m,A}}{\tr \rho^n_\m}\bigg]\;\;\;& = \;\;\; \frac{1}{1-n}\lim_{k \ra 0}\frac{1}{k}\left[\sum_\m p_\m(\tr\rho^n_{\m,A})^k - \sum_\m p_\m(\tr\rho^n_{\m})^k\right] \notag \\
    \label{replicalimit}
    & = \;\;\;\frac{1}{1-n}\lim_{k \ra 0}\frac{d}{dk}\log\left(\frac{\mc Z_A}{\mc Z_0}\right), 
\end{align}
where $\mc Z_A= \sum_\m p_\m(\tr\rho^n_{\m,A})^k$ and $\mc Z_0 = \sum_\m p_\m(\tr\rho^n_{\m})^k$ are the replica partition functions, and the last equation is true because $\lim_{k\ra 0} \mc Z_A = \mc Z_0 =1$ due to the normalization $\sum_\m p_\m = 1$. As discussed in the main text, both $\mc Z_A$ and $\mc Z_0$ admit a Euclidean path integral representation. In particular, setting $Z = \tr \rho = 1$, we identify 
\begin{equation}
\label{pathintrep}
\tr \rho^n_{\m, A} = Z_{\m, \mc M_n} \;\;\mathrm{and}\;\;\; \tr \rho_\m = \lim_{n \to 1} \tr \rho^n_{\m, A} = Z_{\m, \mc M_1} := Z_\m. 
\end{equation}
Here, $Z_{\m, \mc M_n}$ denotes the partition function of the compactified free boson theory on the Riemann surface $\mc M_n$ which contains a branch cut along the region $A$ and boundary conditions $\m$ in region $C$. This leads to
\begin{equation}
\label{zasimplified}
\mc Z_A = \sum_\m Z_\m (Z_{\m, \mc M_n})^k, 
\end{equation}
while $\mc Z_0$ is obtained by first setting $n = 1$ and then replacing $k \to nk$ in the above expression. 

\subsection{Quantum-Classical Split}
We proceed by evaluating the more general object $\mc Z_A$, beginning (as usual) with a conformal map from the Riemann surface $\mc M_n$ to a finite cylinder $\mc C(n)$ of height $h(\crat)/n$ (see equation~\eqref{hxdef}) and circumference $\beta = 2\pi $ such that the boundaries of $\mc M_n$ map to the boundaries of the finite cylinder $\mc C(n)$ (see Fig. 1 of the main text and section \ref{conformalmappingforcedmie}). Through a similar line of reasoning as in \ref{conformalmappingforcedmie}, this results in an analog of \eqref{freeenergysplitforcedmie} which reads
\begin{equation}
    F_{\m, \mc M_n} = F^{\mathrm{geom}}_n + F_{\m, \mc C(n)},
\end{equation}
where we have defined $F_{\m, \mc M_n} := -\log Z_{\m, \mc M_n}$, and $F_{\m, \mc C(n)} := -\log Z_{\m, \mc C(n)}$ is the free energy on the cylinder $\mc C(n)$ with measurement boundary conditions at both boundaries. As discussed in ~\ref{conformalmappingforcedmie} and indicated above, the geometric term $F^{\mathrm{geom}}_n$ depends only on the conformal map and is thus independent of the measurement boundary conditions. Next, we decompose the bosonic field $\vphi$ on $\mc C(n)$ as $\vphi = \vphi_{\mathrm{cl}, \m} + \vphi_q$, separating it into a classical and quantum part. This decomposition leads to a corresponding factorization of the partition function as 
\begin{equation}
\label{explicitqclassicalsplit}
    Z_{\m, \mc C(n)} =Z^q_{D,\mc C(n)}\sum_{\vphi_{{\rm cl},\m}}\exp[-S_{\mc C(n)}[\vphi_{\mathrm{cl,\m}}]], 
\end{equation}
where the former captures the ``quantum contribution" (see \ref{finalexpforcedmie}) and the latter sums over all possible realizations of the classical field $\vphi_{{\rm cl},\m}$. In this case, the different realizations of the classical field configuration $\vphi_{{\rm cl},\m}$ correspond to different winding sectors $w\in \mathbb{Z}$ since the outcome $\m$ is fixed. Therefore, we can replace $\sum_{\vphi_{{\rm cl}}}\ra \sum_{w}$ and including the geometric part, write
\begin{equation}
\label{splitqclassicalgeom}
    Z_{\m, \mc M_n} = \frac{Z^{\mathrm{geom}}_n}{\eta(q_n)} \sum_w\exp\left[-S_{\mc C(n)}[\vphi_{\mathrm{cl}, \m}]\right],
\end{equation}
which corresponds to Equation~(5) of the main text. Here, $Z^q_{D, \mc C(n)} = 1/\eta(q_n)$ captures the ``quantum" contribution (see \ref{finalexpforcedmie}), while the latter is purely classical with $\vphi_{\mathrm{cl}, \m}$ satisfying  the measurement boundary conditions 
\begin{align}
    \vphi_{{\rm cl},\m}\vert_{C_1} &= {\m_1}(\theta)+ 2\pi w, \notag \\
    \vphi_{{\rm cl},\m}\vert_{C_2} &= \m_2(\theta).
\end{align}
Here, $\m_1(\theta)$ and $\m_2(\theta)$ are the respective coarse-grained version of the boundary conditions at the boundaries $C_1$ and $C_2$ with $\theta$ being the polar coordinate on the cylinder, and we sum over all winding sectors $w$ in \eqref{splitqclassicalgeom}. Note that we consider all possible realizations of $\vphi_{\rm cl,\m}$ at the boundaries, and do not restrict to conformally invariant boundary conditions. Next, we remark that \textit{a priori}, both boundaries $C_1$ and $C_2$ have their own set of winding numbers. However, since the free-boson action is invariant under constant shifts of the fields, i.e., $S[\vphi] = S[\vphi+c]$ with $c$ a constant, we can absorb one set of winding numbers, say for $C_2$ as above. Altogether, these steps reduce the evaluation of the otherwise complicated sum over measurement outcomes in~\eqref{zasimplified} to the expression
\begin{equation}
\label{zageompluswindingplusquantum}
    \mc Z_A = \frac{Z^{\mathrm{geom}}_1}{\eta(q_1)} \left( \frac{Z^{\mathrm{geom}}_n}{\eta(q_n)} \right)^k 
    \sum_{\vec{w} \in \mathbb{Z}^{k}} \sum_\m 
    \exp\left[ -\sum_{i=1}^{k} S_{\mc C(n)}[\vphi^{(i)}_{\mathrm{cl}, \m}] 
    - S_{\mc C(1)}[\vphi^{(k+1)}_{\mathrm{cl}, \m}] \right],
\end{equation}
where only the classical contributions enter the summation over measurement outcomes. Additionally, we have a summation over all topological sectors where $\vec{w} = (w_1, \dots, w_{k})$ collects the \textit{relative} winding numbers\footnote{One may, without loss of generality, drop the winding number $w_{k+1}$ for $\vphi^{(k+1)}_{{\rm cl},\m}$ and instead work with the relative winding numbers $ w_i - w_{k+1}$. This is permissible due to the boundary conditions satisfied by the classical fields: $\vphi^{(1)}_{{\rm cl},\m}\vert_{C_1} = \dots = \vphi^{(k)}_{{\rm cl},\m}\vert_{C_1} = {\vphi}^{(k+1)}_{{\rm cl},\m}\vert_{C_1}$. For any pair of replicas, such as $(i, k+1)$ with $i \leq k$, we have $\vphi^{(i)}_{{\rm cl},\m}\vert_{C_1}+2\pi w_i = \vphi^{(k+1)}_{{\rm cl},\m}\vert_{C_1}+2\pi w_{k+1}$ which remains invariant under simultaneous shifts of the form $(w_i, w_{k+1}) \to (w_i + c, w_{k+1} + c)$ for any constant $c \in \mathbb{Z}$. Thus, one may fix this redundancy by choosing a gauge in which $w_{k+1} = 0$.} associated with the set of $k+1$ replicas appearing in~\eqref{zasimplified}. In the above expression, it is clear that the part with the summation over the measurement outcomes provides the key contribution to the MIE, as the remaining terms appear in the forced MIE (see Section~\ref{forcedmiesection}) too, and thus factor out of the measurement sum $\sum_\m$. We now calculate these key classical contributions to the MIE. 

\subsection{Winding Contribution to the MIE}
In the forced measurement case, the absence of a replica structure allowed the classical contribution to take the simplified form~\eqref{classicalpartforced}. In contrast, the classical part of the MIE (see~\eqref{zageompluswindingplusquantum}) involves $k+1$ interacting replicas, which are coupled through the measurement boundary conditions $\m = \m_1 \cup \m_2$ (up to winding). As explained in the main text, the resulting summation over classical actions in $\mc Z_0$ can be simplified by performing a rotation on the $Q = nk + 1$ replicated fields $\bm{\psi} = (\vphi^{(1)}_{{\rm cl},\m},\dots,\vphi^{(Q)}_{{\rm cl},\m})$ to a new basis $\bar{\bm \psi} = (\bar{\vphi}^{(1)}_{{\rm cl},\m},\dots,\bar{\vphi}^{(Q)}_{{\rm cl},\m})$. This technique, first introduced by Fradkin and Moore in the context of 2D quantum critical points~\cite{PhysRevLett.97.050404}, rotates the replica fields such that only a single replica retains an explicit dependence on $\m$, while the remaining $Q-1$ have boundary conditions that vanish (up to winding). Although the original work~\cite{PhysRevLett.97.050404} did not include winding contributions, their relevance was highlighted in subsequent studies~\cite{PhysRevB.79.115421, oshikawa2010boundary, PhysRevLett.107.020402, Zhou_2016}. While our setup differs from those earlier works, similar techniques can be applied here, with appropriate generalizations needed for the evaluation of $\mc Z_A$. $\mc Z_0$, as noted earlier, is then recovered by setting $n=1$ in $\mc Z_A$ and then making the replacement $k \rightarrow nk$.\\

The first step in further simplifying the factor within the measurement summation in \eqref{zageompluswindingplusquantum} is to perform a basis rotation with the aim of isolating a single replica with measurement dependent boundary conditions. There is however, a minor obstruction to directly doing so, namely that the fields being summed over in \eqref{zageompluswindingplusquantum} inhabit different domains: the first $k$ fields reside on $\mc C(n)$, while the final Born-replica field lives on $\mc C(1)$. To resolve this mismatch in the free-boson case, we recast the difference in domains as a difference in boundary conditions—a trade that we can subsequently handle. This is achieved by observing that the action on $\mc C(1)$ can be rewritten as follows:
\begin{equation}
    S_{\mc C(1)}[\vphi_{{\rm cl},\m}] = \frac{g}{4 \pi} \int_0^{\beta = 2 \pi} d \tau \int_0^{h} dy \left(\frac{\delta\m }{ h}\right)^2 = \frac{g}{2}\frac{(\delta \m)^2}{h} = \frac{g}{2}\frac{(\delta \m/\sqrt{n})^2}{h/n} = S_{\mc C(n)}[\tilde{\vphi}_{{\rm cl},\m}], 
\end{equation}
where $\delta\m = \m_1-\m_2+2\pi w$, and $\tilde{\vphi}_{{\rm cl},\m}:=\vphi_{{\rm cl},\m}/\sqrt{ n}$ is a new scaled version of the old field $\vphi_{{\rm cl}}$. Altogether, the boundary conditions for the fields (along with the new scaled field $\tilde{\vphi}_{{\rm cl},\m}$) then read
\begin{align}
        \vphi^{(i)}_{{\rm cl},\m}\vert_{C_1} &= {\m_1}(\theta)+ 2\pi w_i\;\;\;\;i={1,\dots ,k}\\
    \tilde\vphi^{(k+1)}_{{\rm cl},\m}\vert_{C_1} &=\frac{\m_1(\theta)}{\sqrt{n}}, \;\;\;\\
    \vphi^{(i)}_{{\rm cl}}\vert_{C_2} &= \m_2(\theta),\;\;\;\;\;\;\;\;\;\;\;\;\;\;\;\;i = 1,\dots,k\\
    \tilde\vphi^{(k+1)}_{{\rm cl},\m}\vert_{C_2} &= \frac{\m_2(\theta)}{\sqrt{n}}, 
    \label{initialbcs}
\end{align}
where now all the fields $\bm{\vphi}=(\vphi^{(1)}_{{\rm {\rm cl}},\m},\dots,\vphi^{(k)}_{{\rm cl},\m},\tilde{\vphi}^{(k+1)}_{{\rm cl},\m}
)$ live on $\mc C(n)$. Next, we seek a basis rotation of the kind mentioned earlier. The rotation à la Fradkin and Moore~\cite{PhysRevLett.97.050404}, which has been a useful tool in calculating entanglement measures in 2D quantum critical points \cite{PhysRevLett.97.050404, PhysRevLett.107.020402, Zhou_2016, Zhou2_2016, angel-ramelli_entanglement_2019, angel-ramelli_logarithmic_2020, Angel-Ramelli_2021}, fails in this case (when considered in its original form) due to a mismatch in boundary conditions across the replicas (modulo winding). This discrepancy arises entirely from the Born-weighted averaging, which introduces an additional replica with modified boundary conditions in $\mc Z_A$. To resolve this, we are able to construct a more general basis rotation $\mathcal{R}_{k+1,n}$ of size $(k+1)\times (k+1)$ such that $\bar{\bm{\vphi}} = (\bar{\vphi}^{(1)}_{{\rm cl},\m},\dots,\bar{\vphi}^{(k+1)}_{{\rm cl},\m})=\mathcal{R}_{k+1,n}\bm{\vphi}$, and where $\bar{\bm{\vphi}}$ has dependence on $\m$ only on one of its component fields. The transformation $\mc R_{k+1}$ can be thought as reflecting the vector $\vec{\mu} = \left(1,\dots,1/\sqrt{n}\right)^T$ to the vector $\vec{\nu}=||\mu||(0,0,\dots,1)^T$, and is hence the reflection matrix
\begin{equation}
\label{rotationmatrix}
\mathcal{R}_{k+1} = \mathbb{I} - 2\frac{\vec{\gamma}\vec{ \gamma}^T}{\langle\vec{\gamma},\vec{\gamma}\rangle},
\end{equation}
where $\vec{\gamma} =\vec{\mu} - \vec{\nu}$, and $\mc R_{k+1}$ is a unitary transformation by virtue of it being a reflection. Under its action, the boundary conditions transform to
\begin{align}
        \bar\vphi^i_{{\rm cl},\m}\vert_{C_1} &= 2\pi(M_{k})_{ij}w_j\;\;\;\; \rm{for} \;\; i\leq k,
        \label{windingbcs}
        \\
    \bar\vphi^{k+1}_{{\rm cl},\m}\vert_{C_1} &=\sqrt{\frac{nk+1}{n}}\m_1(\theta)+ 2\pi\sqrt{\frac{n}{nk+1}}\sum_{i=1}^kw_i,
    \label{k+1m1}
    \\
    \bar\vphi^i_{{\rm cl},\m}\vert_{C_2} &= 0  \;\;\;\;\;\;\;\;\;\;\;\;\;\;\;\;\;\;\;\;\; \rm{for} \;\;\ i\leq k,
    \label{c2bcs}
    \\
    \bar\vphi^{k+1}_{{\rm cl},\m}\vert_{C_2} &=\sqrt{\frac{nk+1}{n}}{\m_2}(\theta)
    \label{k+1m2},
\end{align}
where $M_k$ denotes the $k\times k$ sub-block of $\mc R_{k+1,n}$ with the $(k+1)$th row and column omitted, and the $\m$ dependence has been relegated only to the final replica as required. Overall, these steps lead to the following simplification for the summation over measurements and winding sectors in \eqref{zageompluswindingplusquantum}:
\begin{equation}
\label{simplifyintroducewinding}
    \sum_{\vec{w} \in \mathbb{Z}^{k}} \sum_\m 
    \exp\left[ -\sum_{i=1}^{k} S_{\mc C(n)}[\vphi^{(i)}_{\mathrm{cl}, \m}] 
    - S_{\mc C(1)}[\vphi^{(k+1)}_{\mathrm{cl}, \m}] \right] = \mc W_{n,k}\int {\mathcal D}[\m(\theta)] \exp[-S_{\mc C(n)}[\bar{\vphi}^{(k+1)}_{{\rm cl},\m}]], 
\end{equation}
where we have written $\sum_\m$ as a functional integral $\int {\mathcal D}[\m(\theta)]=\int {\mathcal D}[\m_1(\theta)]{\mathcal D}[\m_2(\theta)]$ over the coarse-grained outcomes $\m(\theta)$, and have defined
\begin{equation}
\label{windingdef}
    \mc W_{n,k}: = \sum_{\vec{w}\in\mathbb{Z}^k}\exp\left[-\sum_{i=1}^k S_{\mc C(n)}[\bar{\vphi}^{(i)}_{{\rm cl}}]\right],
\end{equation}
as the ``winding function" (see~\cite{Zhou_2016} for other examples) that is independent of measurement outcomes, allowing it to be pulled out of the summation $\sum_\m$. We evaluate the summation over measurements for this single replica and return to the winding function later. We do this by first noting that the transformation \eqref{rotationmatrix} changes the compactification radius $r$ of  $\bar{\vphi}^{(k+1)}_{{\rm cl},\m}$ from $r=1$—which is the convention we have followed throughout this work—to $\sqrt{(nk+1)/n}$. In order to return to the usual conventions of $r=1$, we define a re-scaled field $\vphi'^{(k+1)}_{{\rm cl},\m} = \sqrt{n/(nk+1)}\bar{\vphi}_{{\rm cl},\m}^{(k+1)}$ which has radius of compactification $r=1$. Note that we perform this re-scaling only to have a meaningful Luttinger parameter $g$ which is defined in the main text as the free parameter in the TLL setup with a fixed $r=1$. We can then write \eqref{simplifyintroducewinding} purely in terms of the re-scaled field, giving
\begin{equation}
    \int {\mathcal D}[\m(\theta)] \exp[-S_{\mc C(n)}[\bar{\vphi}^{(k+1)}_{{\rm cl},\m}]] = \int {\mathcal D}[\vphi'^{(k+1)}_{{\rm cl},\m}|_C] \exp\left[-S_{\mc C(n)} \left[\sqrt{\frac{nk+1}{n}}\vphi'^{(k+1)}_{{\rm cl},\m} \right]\right],
\end{equation}
where we have used ${\mathcal D}[\m(\theta)] = {\mathcal D}[\vphi'^{(k+1)}_{{\rm cl},\m}|_C]$. Here, the factor of $\sqrt{(nk+1)/n}$ can now be absorbed by defining a renormalized Luttinger parameter $((nk+1)/n)g$ where the compactification radius of $\vphi'^{(k+1)}_{{\rm cl},\m}$ is unity. Doing this, gives 
\begin{equation}
\label{randomeqn}
    \int {\mathcal D}[\m(\theta)] \exp[-S_{\mc C(n)}[\bar{\vphi}^{(k+1)}_{{\rm cl},\m}]] = \int {\mathcal D}[\vphi'^{k+1}_{{\rm cl},\m}|_C] \exp\left[-S_{\mc C(n),\frac{nk+1}{n}g}\left[\vphi'^{(k+1)}_{{\rm cl},\m}\right]\right],
\end{equation}
where we have made the Luttinger parameter dependence in the action explicit. Finally, right-hand side above can be re-written in terms of standard functions using the completeness relation
\begin{equation}
Z^n= (\tr\rho)^n= 1 = \sum_{\m} Z_{\m, \mc C(n)} =Z^q_{D,\mc C(n)}\int {\mathcal D}\left[\bar{\vphi}'^{(k+1)}_{{\rm cl},\m}\right]\exp[-S_{\mc C(n),g}[\bar{\vphi}'^{(k+1)}_{{\rm cl},\m}]],
\end{equation}
where $Z^q_{D,\mc C(n)} = 1/\eta(q_n)$, and so \eqref{randomeqn} is $g$-independent. Using this, we can simplify \eqref{simplifyintroducewinding} and in turn \eqref{zageompluswindingplusquantum} as
\begin{equation}
\label{zasemifinal}
    \mc Z_A = \mc W_{n,k} \frac{Z^{\mathrm{geom}}_1}{\eta(q_1)} \left( \frac{Z^{\mathrm{geom}}_n}{\eta(q_n)} \right)^k \eta(q_n).
\end{equation}
To summarize, we have simplified $\mc Z_A$ from its form in \eqref{zageompluswindingplusquantum} by capturing the effect of interactions between the $k+1$ replica fields into the winding function $\mc W_{n,k}$, while leaving behind a \textit{single} free-replica $\bar{\vphi}^{(k+1)}_{{\rm cl},\m}$ that we can sum over to give $(Z^{q}_{D,\mc C(n)})^{-1}=\eta(q_n)$. Alternatively, the winding function accounts for compactification in the new basis $\bar{\bm \vphi}$ which arises precisely because of the interacting replicas in~\eqref{zageompluswindingplusquantum}. The form of this basis transformation is in turn governed entirely by the replica structure of $\mc Z_A$ which includes $k$ replicas arising from $\tr \rho^n_{\m, A}$ and a single Born-replica coming from the weight $\tr\rho_\m = p_{\m}$ (see~\eqref{pathintrep} and~\eqref{zasimplified}). 

\subsection{Evaluation and Analytic Continuation of $\mc W_{n,k}$}
We now present an exact evaluation of the winding contribution $\mc W_{n,k}$ which—as discussed in the main text—is central to the MIE and crucial in distinguishing it from $\mathrm{MIE}_\mathrm{F}$. The evaluation and subsequent analytic continuation of $\mc W_{n,k}$ in $k$ discussed in this section closely follows that of a very similar winding function obtained in Ref.~\cite{Zhou_2016} in a different context. 

We start by performing a similar re-scaling of the fields $\bar{\vphi}'^{(i)}_{{\rm cl},\m}= ((M_k^T\mathbf 1_k)_i)^{-1}\bar{\vphi}^{(i)}_{{\rm cl},\m},\;i=1,\dots,k$ as done in the previous section to return to our conventions of $r=1$. Here, $\mathbf{1}_k = (1,1,\dots,1)^T$ is a $k\times 1$ sized matrix of ones. Again, this is because the transformation $\mc R_{k+1}$ changes the compactification radius of the fields $\bar{\vphi}^{(i)}_{{\rm cl},\m}$ from $r=1$ to $r=(M_k^T\mathbf 1_k)_i$ as seen by \eqref{windingbcs}. The re-scaled fields now all have $r_i=1$ and $g_i = ((M_k^T\mathbf 1_k)_i)^2g$. Again, this re-scaling step is purely done to assign renormalized Luttinger parameters to the rotated fields $\{\bar{\vphi}^{(i)}_{{\rm cl},\m}\},\;i=1,\dots,k$ and is not essential as far as the mathematical form of the final expression goes which depends on the combination $g_ir_i^2$. Next, we expand out the actions in \eqref{windingdef} as
\begin{equation}
    \mc W_{n,k} = \sum_{\vec{w}\in\mathbb{Z}^k}\exp\left[-\sum_{i=1}^{k}\frac{g_i}{4 \pi} \int_0^{\beta = 2 \pi} d \tau \int_0^{h/n} dy \frac{(2\pi w_i)^2}{ (h/n)^2}\right] = \sum_{\vec{w}\in \mathbb{Z}^{k}}\exp\bigg[-\frac{gn}{2h}(2\pi)^2\vec{w}^T M_k^TM_k \vec{w}\bigg] =  \sum_{\vec{w}\in \mathbb{Z}^{k}}q_n^{g\vec{w}^TT_k\vec{w}},
\end{equation}
where we have defined $T_k = M_k^TM_k$. The above has a similar structure as the classical part \eqref{classicalpartforced} of the forced case. This is not surprising because the classical part simply sums over different winding sectors like \eqref{classicalpartforced} but for a set of $k-$replicated fields, all with different Luttinger parameters (or equivalently, compactification radii). Next, in order to analytically continue the above, it is convenience to perform apply the reciprocal formula to the above expression \cite{7a6ba7cf-6d4a-3566-be69-751572b726ce} which gives
\begin{equation}
\label{poissonresummedwn}
\mc W_{n,k}=\sum_{\vec{w}\in \mathbb{Z}^{k}}q_n^{g\vec{w}^TT_k\vec{w}}= \frac{\sqrt{nk+1}}{(2\pi ng/h)^{k/2}}\sum_{w\in\mathbb{Z}^k}\tilde{q}_n^{\vec{w}^TT^{-1}_k\vec{w}/4g}, 
\end{equation}
where $\tilde q_n = e^{-\frac{4\pi}{\beta}\frac{h(\crat)}{n}} = e^{-2h(\crat)/n}$ and 
\begin{equation}T^{-1}_k =  \begin{bmatrix}
    n+1& n &n & \dots&\dots&\dots&\dots \\
n & n+1 &n&\dots&\dots&\dots&\dots \\
n &n &n+1 & \dots &\dots&\dots&\dots\\
\dots & \dots & \dots & \dots &\dots&\dots&\dots \\ 
\dots & \dots &\dots &\dots&\dots & n+1 & n\\
\dots & \dots & \dots & \dots &\dots & n & n+1
\end{bmatrix}_{k\times k}
\end{equation}
is now a $k-$independent matrix. Analytic continuation proceeds by completing the square in $\vec{w}^TT^{-1}_k\vec{w}$  \cite{Zhou_2016}
\begin{equation}
    \vec{w}^TT^{-1}_k\vec{w} = n\left(\sum_{i=1}^k w_i\right)^2 + \sum_{i=1}^k w_i^2,
\end{equation}
and re-writing \eqref{poissonresummedwn} as an integral with a Kronecker delta, returning 
\begin{align}
\sum_{w\in\mathbb{Z}^k}\tilde{q}_n^{\vec{w}^TT^{-1}_k\vec{w}/4g} &= \sum_{\vec w\in\mathbb{Z}^k}\exp\left[-\frac{h}{2ng}\sum_{i=1}^k w_i^2 - \frac{h}{2g}\left(\sum_{i=1}^k w_i\right)^2\right] \notag \\
&=  \int_0^{2\pi}\frac{d\delta_\vphi}{2\pi}\sum_{\vec w\in\mathbb{Z}^k}\sum_{l\in \mathbb Z} \exp\left[-\frac{h}{2ng}\sum_{i=1}^k w_i^2 - \frac{h}{2g}l^2\right]\exp\left[i\delta_\vphi\left(l-\sum_i w_i\right)\right].
\end{align}
One can factorize the sum over $\vec w$ into $k$ independent sums, resulting in
\begin{equation}
\sum_{w\in\mathbb{Z}^k}\tilde{q}_n^{\vec{w}^TT^{-1}_k\vec{w}/4g} =  \int_0^{2\pi}\frac{d\delta_\vphi}{2\pi} \sum_{l\in\mathbb Z}\exp\left[-\frac{h}{2g}l^2+il\delta_\vphi\right]\left[\sum_{w\in\mathbb Z}\exp\left(-\frac{h}{2ng}w^2-i\delta_\vphi w\right)\right]^k.
\end{equation}
Using the Poisson-summation formula
\begin{equation}
    \sum_{w\in\mathbb{Z}} \exp\bigg[-\pi a w^2 + 2\pi i b w\bigg] = \frac{1}{\sqrt{a}}\sum_{w\in \mathbb{Z}}\exp\bigg[ - \frac{\pi}{a}(w+b)^2\bigg],
\end{equation}
one can re-write the sum over $w$ and $l$, giving
\begin{equation}
   \sum_{w\in\mathbb{Z}^k}\tilde{q}_n^{\vec{w}^TT^{-1}_k\vec{w}/4g}= \left(\frac{2\pi ng}{h}\right)^{k/2}\left(\frac{2\pi g}{h}\right)^{1/2}\int_0^{2\pi}\frac{d\delta_\vphi}{2\pi}\sum_{l\in\mathbb Z}\exp\left[-\frac{2\pi^2 g}{h}\left(l+\frac{\delta_{\vphi}}{2\pi}\right)^2\right] \left[\sum_{w\in\mathbb Z}\exp\left(-\frac{2ng\pi^2}{h}\left(w+\frac{\delta_\vphi}{2\pi}\right)^2\right)\right]^k.
\end{equation}
We can write the above in a more suggestive form
\begin{equation}
\label{windingcontributionza}
    \mc W_{n,k} = \sqrt{\frac{ Qg}{2\pi h}}\int_0^{2\pi} d\delta_\vphi\sum_{l\in\mathbb{Z}}e^{-\frac{2\pi^2g}{h}\left(l+\frac{\delta_\vphi}{2\pi}\right)^2}\left[\sum_{w\in\mathbb Z}q_n^{g(w+\delta_\vphi/2\pi)^2}\right]^k,
\end{equation}
where we have written the sum over $w$ using $q_n = e^{-2\pi^2 ng/h}$. This equation has a natural interpretation as an average over boundary conditions over all replicas with the distribution $P(\delta_\vphi) = \sqrt{\frac{g}{2\pi h}} \sum_{l\in\mathbb{Z}}e^{-\frac{2\pi^2g}{h}\left(l+\frac{\delta_\vphi}{2\pi}\right)^2} $, since $\sum_{w\in\mathbb Z}q_n^{g(w+\delta_\vphi/2\pi)^2}$ is the classical (winding) contribution to the cylinder partition function of a free boson  with boundary conditions $(\vphi_1,\vphi_2)$ at $(C_1,C_2)$ indexed by $\delta_{\vphi}=\vphi_1-\vphi_2$. We can also write
\begin{equation}
\label{windingcontributionzb}
    \mc W_{n,k} = \sqrt{\frac{Qg}{2 \pi h}}\int_{-\infty}^{\infty} d\delta_\vphi e^{- \frac{g\delta_{\vphi}^2}{2h}} \left[\sum_{w\in\mathbb Z}q_n^{g(w+\delta_\vphi/2\pi)^2}\right]^k, 
\end{equation}
which is equation 6 in the main text. The above form can be obtained by breaking up \eqref{windingcontributionza} into infinitely many pieces of $2\pi$ length, covering the full real line. 

\subsection{Replica Limit}
We now take the replica limit exactly to obtain a closed-form expression for the MIE. Using equations~\eqref{zasemifinal}, we write $\mc Z_0$ by first setting $n = 1$ and subsequently replacing $k \to nk$, yielding
\begin{equation}
\mc Z_0 = \mc W_{1,nk} \left( \frac{Z^{\mathrm{geom}}_1}{\eta(q_1)} \right)^{Q} \eta(q_1).
\end{equation}
When considering the $k-$dependent terms in $\log \mc Z_A/\mc Z_0$, we find that the geometric contribution appears as $\log Z^{\mathrm{geom}}_n-n\log Z^{\mathrm{geom}}_1=0$ which, by the arguments presented in section~\ref{conformalmappingforcedmie}, evaluates to 0. Using the replica limit, the MIE can then be written as
\begin{equation}
\label{finaleqmien}
\mathrm{MIE}^{(n)}(A) = \frac{1}{1-n}\lim_{k \ra 0}\frac{d}{dk}\log\left(\frac{\mc Z_A}{\mc Z_0}\right) = \frac{1}{1-n}\left[\mc W_n'-n\mc W_1' +n\log\eta(q_1)-\log\eta(q_n)\right],
\end{equation}
where $\mc W_n' := \lim_{k\ra 0}\pd k \mc W'_{n,k}$ is
\begin{equation}
\label{eqnwnprime}
    \mc W_n' =\sqrt{\frac{g}{2\pi h}} \int_{-\infty}^{\infty}d \delta_{\vphi}\;e^{-\frac{g\delta_{\vphi}^2}{2h}}\log\sum_{w\in\mathbb Z}q_n^{g\left(w+\frac{\delta_{\vphi}}{2\pi}\right)^2 }+\frac{n}{2}.
\end{equation}
Equation~\eqref{finaleqmien} is equation 7 from the main text and is a key result of this work. Similar to $\mief$, the MIE is conformally invariant with the dependence of cross ratio being encoded in $h(\crat)$ despite the presence of boundary conditions (measurement outcomes) that break conformal invariance. Furthermore, the analytic expression \eqref{windingcontributionzb} for the winding function $\mc W_{n,k}$ allows us to take the replica limit in both $k$ and $n$ exactly, thereby leading to a closed form expression for the MIE for all Rényi indices. 

\subsection{Born-averaging over Dirichlet boundary conditions interpretation}

Our results have a natural physical interpretation in terms of Born-averaging over Dirichlet (conformal) boundary conditions that we now make more transparent. First, we note that the probability distribution $P(\delta_\vphi) = \sqrt{\frac{g}{2\pi h}} \sum_{l\in\mathbb{Z}}e^{-\frac{2\pi^2g}{h}\left(l+\frac{\delta_\vphi}{2\pi}\right)^2} $ is directly proportional to the partition function 
\begin{equation}
Z_{\mc C(1),\delta_\vphi} = \frac{1}{\eta(q_1)} \sum_{w\in\mathbb{Z}} q_1^{g \left(w+\frac{\delta_\vphi}{2\pi}\right)^2}, 
\end{equation}
of the compact free boson on the cylinder, with Dirichlet boundary conditions $(\vphi_1,\vphi_2)$ at $(C_1,C_2)$ indexed by $\delta_{\vphi}=\vphi_1-\vphi_2 \in [0, 2\pi)$. In terms of this quantity, we have:
\begin{equation}
\mathcal{Z}_0 = \sum_{\bf m} p_{\bf m}^{nk +1} \sim \int_0^{2 \pi} d \delta_\vphi (Z_{\mc C(1),\delta_\vphi} )^{nk+1},
\end{equation}
where we have focused on the winding contribution and dropped unimportant factors for clarity (including the geometric factor for example). Our results then show that $\mathcal{Z}_A$ has a similarly intuitive form
\begin{equation}
\mathcal{Z}_A = \sum_{\bf m} p_{\bf m} \left(\text{tr} \rho_{{\bf m},A}^n\right)^k \sim \int_0^{2 \pi} d \delta_\vphi Z_{\mc C(1),\delta_\vphi}  (Z_{\mc C(n),\delta_\vphi} )^{k},
\end{equation}
where $Z_{\mc C(n),\delta_\vphi} = \frac{1}{\eta(q_n)} \sum_{w\in\mathbb{Z}} q_n^{g \left(w+\frac{\delta_\vphi}{2\pi}\right)^2} $ follows from replacing $q_1$ by $q_n$. We see that the variable $\delta_\vphi$ directly labels measurement outcomes, with the associated Born probability $\sim Z_{\mc C(1),\delta_\vphi}$. As a result, if we denote by $\miefn(\delta_\varphi)$ the MIE obtained by forcing Dirichlet (uniform) measurement outcomes $(\vphi_1,\vphi_2)$ at $(C_1,C_2)$ (with $\delta_{\vphi}=\vphi_1-\vphi_2$), we find that the MIE corresponding to real measurements is simply given by Born averaging
\begin{equation}
\mathrm{MIE}^{(n)} = \int_0^{2\pi} d\delta_{\vphi}\;P(\delta_\vphi) \miefn(\delta_\varphi), 
\end{equation}
where again, $P(\delta_\vphi) \sim Z_{\mc C(1),\delta_\vphi}$, and
\begin{equation}
\miefn(\delta_\vphi) = \frac{1}{1-n} \log\frac{Z_{\mc C(n),\delta_\vphi}}{Z_{\mc C(1),\delta_\vphi}^n}.
\end{equation}

\subsection{Small cross ratio asymptotics of the MIE}
In this section, we derive the small cross-ratio ($\crat \to 0$) behavior of the MIE—the regime in which most of the system is measured, resulting in a sudden measurable change in the entanglement structure with novel universal behavior. Similar to the forced case, the leading contribution in the small $\crat$ limit comes from the winding terms
\begin{equation}
\label{mieleading}
   \mathrm{MIE}^{(n)}(A) \underset{{\crat\ra 0} }{\sim} \frac{1}{1-n}(\mc W_n' -n\mc W'_1).
\end{equation}
However, the leading behavior of the MIE differs starkly to the forced case \cite{PhysRevB.109.195128} as we now show. For this we need to first extract the behavior of $\mc W_n'$ near $\crat=0$. We start by pulling out the $w=0$ term from the logarithm in \eqref{eqnwnprime}. This piece cancels exactly with the $n/2$ leaving us with
\begin{equation}
    \mc W_n' =\sqrt{\frac{g}{2\pi h}} \int_{-\infty}^{\infty}d \delta_{\vphi}\;e^{-\frac{g\delta_{\vphi}^2}{2h}}\log\sum_{w\in\mathbb Z}q_n^{g\left(w^2+w\delta_{\vphi}/\pi\right) }.
\end{equation}
For $\crat\ra 0$, $h(\crat) = \pi^2/\log(16/\crat) + {\cal O}(\crat/\log^2(\crat))$ and $q_n = (\crat/16)^{2n}$ are small. Therefore we can approximate $\mc W_n'$ by taking only the $w=1,-1$ terms. This returns
\begin{equation}
\label{asymptoticwprimen}
    \mc W_n' = \sqrt{\frac{g}{2\pi h}}\int_{-\infty}^{\infty}d\delta_\vphi \exp\left(\frac{-g\delta_\vphi^2}{2h}\right)\log\left[1+\exp\left(-\frac{2\pi^2ng}{h}\left(1+\frac{\delta_{\vphi}}{\pi}\right)\right)+\exp\left(-\frac{2\pi^2ng}{h}\left(1-\frac{\delta_{\vphi}}{\pi}\right)\right)\right]. 
\end{equation}
We divide the integral into three regions: $(-\infty, -\pi)$, $(-\pi, \pi)$, and $(\pi, \infty)$, corresponding to different dominant contributions to the integrand. Denoting the integrals over these regions by $I_1$, $I_2$, and $I_3$ respectively, we write $\mc W_n' = I_1 + I_2 + I_3=2I_1 + I_2$, where the last equation follows from $\mc W_n'$ being even. For the integral $I_1$, one of the exponentials inside the logarithm dominates. Accordingly, expanding in $h$ as our small parameter, we have
\begin{equation}
    I_1 \approx  -\sqrt{\frac{g}{2\pi h}}\int_{-\infty}^{-\pi}d\delta_\vphi \exp\left(\frac{-g\delta_\vphi^2}{2h}\right)\left[\frac{2\pi^2 ng}{h}\left(1+\frac{\vphi}{\pi}\right)\right] = n\exp\left(\frac{-g\pi^2}{2h}\right)\sqrt{\frac{2\pi g}{h}} - \frac{gn\pi^2}{h}\mathrm{Erfc}\left[\pi\sqrt{\frac{g}{2h}}\right],
\end{equation}
where $\mathrm{Erfc}(z)=2\pi^{-1/2}\int_0^z e^{-t^2}\;dt $ is the error function. We do not further expand the above integral since it is linear in $n$ and does not contribute in \eqref{mieleading}. In $I_2$, both the exponentials inside the logarithm in \eqref{asymptoticwprimen} are much smaller than~$1$. Using $\log(1+\epsilon)\approx \epsilon$, we can write $I_2$ as 
\begin{align}
    I_2 &\approx  \sqrt{\frac{g}{2\pi h}}\int_{-\pi}^{\pi}d\delta_\vphi \exp\left(\frac{-g\delta_\vphi^2}{2h}\right)\left[\exp\left(-\frac{2\pi^2ng}{h}\left(1+\frac{\delta_{\vphi}}{\pi}\right)\right)+\exp\left(-\frac{2\pi^2ng}{h}\left(1-\frac{\delta_{\vphi}}{\pi}\right)\right)\right] \notag \\
    &  = \exp\left(-\frac{2\pi^2g}{h} n(1-n)\right)\left[\mathrm{Erfc}\left[\pi\sqrt{\frac{ g}{2h}}\left(1-2n\right)\right]+\mathrm{Erfc}\left[\pi\sqrt{\frac{ g}{2h}}\left(1+2n\right)\right]\right] \notag \\
    & = \left(\mathrm{sgn}\left(1-2n\right)+\mathrm{sgn}(1+2n) \right)\exp\left(-\frac{2\pi^2g}{h} n(1-n)\right) - \sqrt{\frac{h}{2\pi g}}\frac{2}{\pi(1-2n)}\exp\left(\frac{-\pi^2 g}{2h}\right)+{\cal O}\left(h^{3/2}e^{\frac{-\pi^2 g}{2h}}\right).
\end{align}
The above expansion indicates a transition in MIE scaling at $n=1/2$. Extracting the leading behavior of the MIE using $I_2$ gives
\begin{equation}
    \mathrm{MIE}^{(n)}(A) \approx \begin{cases}
\displaystyle 
\frac{2n+1}{2n-1}\frac{1}{\pi}\sqrt{\frac{2h}{\pi g}}e^{-\frac{\pi^2g}{2h}}
& n > \tfrac12,\\[1ex]
\displaystyle
\,e^{-\frac{\pi^2g}{2h}}
& n = \tfrac12, \\[1ex]
\displaystyle
2\,e^{-\frac{\pi^2g}{2h}\,n(1-n)}
& 0 < n < \tfrac12.
\end{cases}
\end{equation}
Using $h(\crat) =\pi^2/\log(16/\crat) + {\cal O}(\crat/\log^2(\crat)) $, we get
\begin{equation}
    \mathrm{MIE}^{(n)}(A) \underset{\crat \to 0}{\sim} \begin{cases}
\displaystyle 
\frac{\crat^{g/2}}{\sqrt{\log(1/\crat)}}
& n > \tfrac12,\\[1ex]
\displaystyle
\,\crat^{g/2}
& n = \tfrac12, \\[1ex]
\displaystyle
\crat^{2gn(1-n)}
& 0 < n < \tfrac12,
\end{cases}
\end{equation}
where we dropped unimportant prefactors. 
The leading behavior of the MIE displays distinctive features, particularly when contrasted with the forced case (see Section~\ref{forcedmiexll1}). First, it is now confirmed that for $n>1/2$ the leading exponent for the MIE ($g/2$) is indeed smaller than that of the forced case ($2g$) as was already seen numerically in Ref.~\cite{PhysRevB.109.195128}. In the same regime, a multiplicative factor $1/\sqrt{\log(1/\crat)}$ appears—an unexpected feature given that standard operator product expansion (OPE) analyses of CFT entanglement entropies have previously only reported integer power logarithms~\cite{rajabpour2012entanglement, ruggiero2018entanglement}. At $n=1/2$, the scaling behavior undergoes a qualitative change. Below $n=1/2$, we see a quadratic $n$-dependent scaling, in contrast to the forced case where the scaling grows linearly in $n$ up until $n=1$ and then saturates to $2g$. 

Finally, as discussed in the main text, MIE includes pre-existing entanglement. A related but more fitting quantity that isolates the entanglement induced due to measurements is the measuremenet-induced information (MII) defined in Ref.~\cite{PhysRevB.109.195128} as
\begin{equation}
\label{MIIdef}
    \mathrm{MII}(A, B) = \sum_\m p_\m I(A,B)[\ket{\psi_\m}] - I(A,B)[\ket{\psi}],
\end{equation}
where $I(A, B)$ is the mutual information between $A$ and $B$, defined as
\begin{equation}
    I(A,B)= S(A) + S(B) - S(AB).
\end{equation}
In our setup, it is easy to show that $\sum_\m p_\m I(A,B)[\ket{\psi_\m}] = 2\mathrm{MIE(A,B)}$ using the fact that the full state is pure. Therefore, we can use the leading MIE behavior above to predict how MII behaves in the regime of small $\crat$ where the system undergoes sudden change in entanglement structure. Restricting to the case of free-fermions with $g=1/2$, we see that the forced MIE scales as $\sim \crat$ and the MIE scales as $\crat^{1/4}$ with a logarithmic factor. Furthermore, 1+1D CFT calculations show that mutual information scales as $\sim \crat^{1/2}$ in the small $\crat$ limit~\cite{Calabrese_2009_disjoint}. Thus, using \eqref{MIIdef} and the fact that $\sum_\m p_\m I(A,B)[\ket{\psi_\m}] = 2\mathrm{MIE(A,B)}$, it is clear that the MII is actually negative for the forced case (where we post-select to the anti-ferromagnetic state), implying that such a forced measurement \textit{decreases} correlations in the un-measured system. On the other hand, real measurements have a positive MII that \textit{increase} correlations.

\section{Numerics}
In order to further benchmark our results, we compare them to exact and matrix product states (MPS) calculations on the XXZ-chain. At $\Delta=0$, we map the model to free-fermions at half-filling with the Hamiltonian
\begin{equation}
    H = -\sum_i \left(c_i^{\dag}c_{i+1} +\; \text{h.c.}\right)+\mathrm{const.}
\end{equation}
with periodic (antiperiodic) boundary conditions when the total number of fermions is odd (even). The numerics for this case can be done exactly since the ground state is Gaussian and hence is entirely determined by its correlation matrix \cite{giamarchi_quantum_2003}
\begin{equation}
    C_{ij}=  \langle c_i^{\dag}c_j\rangle = \frac{\sin(\pi n_f(i-j))}{L\sin\frac{\pi(i-j)}{L}},
\end{equation}
where $n_f$ is the fermion-filling factor which is $1/2$ in our case. $\sigma_z$ basis measurements are then made by updating the correlation matrix as per the update rules
% --- Eq. (A3)
\begin{equation}
C'_{ij}=\frac{\langle c_a^\dagger c_a\, c_i^\dagger c_j\, c_a^\dagger c_a\rangle}{C_{aa}}
=\begin{cases}
1, & i=j=a,\\[4pt]
C_{ij}-\dfrac{C_{ia}C_{aj}}{C_{aa}}, & i\neq a,\; j\neq a,\\[8pt]
0, & \text{otherwise},
\end{cases}
\tag{A3}
\end{equation}
when we apply the projector $P_1 = c^{\dag}_{a}c_a$ with probability $p_a = C_{aa}$, where $a$ is the measured orbital (site). Similarly, when we apply the projector $P_0 = 1-c_a^{\dag}c_a$ with probability $p_0=1-C_{aa}$, the updated correlation matrix is
% --- Eq. (A4)
\begin{equation}
C'_{ij}=\frac{\langle c_a c_a^\dagger\, c_i^\dagger c_j\, c_a c_a^\dagger\rangle}{1-C_{aa}}
=\begin{cases}
0, & i=j=a,\\[4pt]
C_{ij}+\dfrac{C_{ia}C_{aj}}{1-C_{aa}}, & i\neq a,\; j\neq a,\\[8pt]
0, & \text{otherwise}.
\end{cases}
\tag{A4}
\end{equation}
where multi-particle correlators can be evaluated using Wick’s theorem. The above rules are easy to derive (see Ref.~\cite{PhysRevB.109.195128}). In the forced case, we post-select to the anti-ferromagnetic state $\ket{1010\dots}$ whereas for the MIE, we sample measurements as per the Born rule. This is done as follows: for each measured site $i$, we first consider the updated correlation matrix associated with applying the projector $P^{(i)}_0$ and compute its Born probability $p^{(i)}_0=\langle P^{(i)}_0\rangle = 1-\langle c_i^\dag c_i\rangle = 1-C_{ii}$. We then draw a random number $r$ from the uniform distribution $\mc U[0,1]$. If $r<p^{(i)}_0$, we accept this outcome and keep the corresponding updated correlation matrix. Otherwise, we discard this temporary update and instead apply the orthogonal projector $P^{(i)}_1$, updating the pre-measurement correlation matrix. Iterating this procedure over all measured sites $i$ generates measurement outcomes distributed according to the Born rule.
\\ 
Post-measurement entanglement entropy is then easily calculated from the correlation matrix of the part of the system that has not been measured~\cite{Peschel_2009}. We display the numerical results for system sizes $L = 200,300$ and $400$ and Rényi indices $n=0.5,1.0$ and $3.0$ for the forced MIE in Fig.\ref{mieforcedfig}. For the same Rényi indices, we plot the MIE in the main text with $L=300,400$ and $600$. The number of samples in the plot depend on the cross ratio $\crat$ and range from $4\times 10^3$ to $2\times 10^4$ samples. 

For $\Delta\neq0$, we use the \texttt{iTensor} library \cite{itensor} to obtain approximate ground states using DMRG \cite{white1992density, white1993density}. Measurements are made by applying standard projection operators to the MPS in the $\sigma_z$ basis. We perform DMRG simulations for system sizes $L = 120$, $160$, and $200$, with interaction strengths $\Delta = -0.3$ and $0.5$. The bond dimension is increased during the sweeps, subject to a maximum truncation error threshold of $10^{-7}$. The resulting maximum bond dimensions are: (364, 312) for $L = 120$, (445, 379) for $L = 160$, and (512, 436) for $L = 200$, where each pair refers to $\Delta = (0.5, -0.3)$, respectively. The plots for MIE in the main text have $3\times 10^3$ to $9\times 10^3$ samples, depending on the value of cross-ratio $\crat$. 

In all of the above cases, the cross-ratios are sampled by fixing $x_1 = 1, x_3 = L/2$ and varying $x_2$ and $x_4$ such that $x_{12} = x_{34}$. This way, the measured regions are always placed antipodally and have same sizes, eliminating any spurious odd-even effects.

\bibliography{apssamp}% Produces the bibliography via BibTeX.